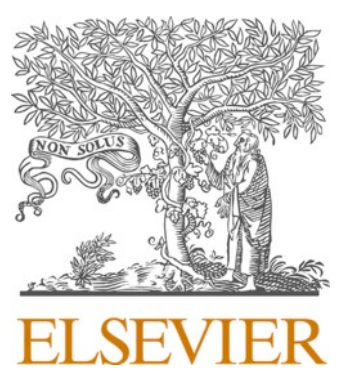

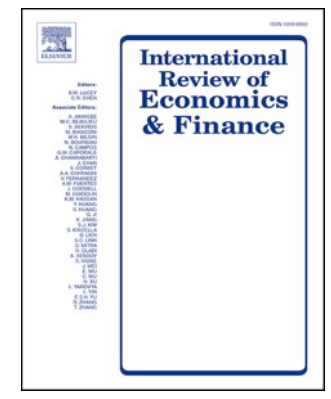

# Prediction of bank transaction fraud using TabNet—an adaptive deep learning architecture

B.S. Prashanth [a], Manoj Kumar [a], Ariful Hoque [b,*], Nasser Al Muraqab [c], Immanuel Azaad Moonesar [d,e], Udo Christian Braendle [f], Ananth Rao [g]

[a] Department of Information Science and Engineering, Nitte Meenakshi Institute of Technology, Bengaluru, India
[b] Finance, Murdoch University, Australia
[c] Information Systems, Dubai Business School, University of Dubai, Dubai, United Arab Emirates
[d] Department of Academic Affairs - Public Health, Policy, Systems Research, and, Scientific Policy, International Vaccine Institute, Seoul, South Korea
[e] Mohammed Bin Rashid School of Government (MBRSG), Dubai, United Arab Emirates
[f] CEO and University Management Research & Innovation, IMC Krems University of Applied Sciences, A-3500, Krems, Republic of Austria
[g] Finance, University of Dubai, Dubai Business School, Mohammed Bin Rashid School of Government, Dubai, United Arab Emirates

ARTICLE INFO

ABSTRACT

The development of online banking has brought about an increase in fraudulent operations, which is a major problem for banks. This study delves into the urgent requirement for interpretable, scalable, and top-notch fraud detection systems by using TabNet, an adaptable deep learning framework, on a Kaggle dataset consisting of actual bank transactions in India. Maximizing operational risk management by improving the accuracy of transaction anomaly detection and ensuring regulatory compliance through transparent models is the goal.

We utilize a supervised learning pipeline that incorporates the Synthetic Minority Oversampling Technique (SMOTE) to ensure that classes are balanced. Subsequently, we conduct thorough exploratory data analysis (EDA) to identify patterns of fraud, both during specific times and across behaviors. On this dataset, five different deep learning architectures are tested: DNN, GRU, LSTM, CNN1D, and TabNet. Assessment of predictive performance was carried out using a 3-fold cross-validation framework. With a ROC-AUC of 0.9739 and an accuracy of 97.39 %, TabNet considerably outperformed the competition. The method of sparse feature selection used improved interpretability, generalized better on tabular data, and produced fewer false positives and negatives.

Critical insights for operational fraud detection systems and a contribution to the broader literature on explainable AI (XAI) in financial decision-making are offered by the findings. Goals 8 and 16 of the Sustainable Development Agenda are supported by this study, which promotes inclusive economic growth and institutional transparency. Supporting strong, policy-compliant, and interpretable decision-support systems, it also offers practical use for real-time implementation in banking infrastructure.

* Corresponding author. Murdoch Business School, Murdoch University, Murdoch University, 90 South Street, Murdoch, WA, 6150, Australia.
*E-mail addresses:* prashanth.bshivanna@gmail.com (B.S. Prashanth), manojmv24@gmail.com (M. Kumar), A.Hoque@murdoch.edu.au (A. Hoque), nasser@ud.ac.ae (N. Al Muraqab), immanuel.moonesar@mbrsg.ac.ae (I.A. Moonesar), udo.braendle@imc.ac.at (U.C. Braendle), arao@ud.ac.ae (A. Rao).

## 1. Introduction

Electronic transactions have transformed financial operations. As a result, cybercrime has proliferated, targeting e-banking. Worldwide, fraud costs businesses $4 trillion, or 5 % of their sales. This includes money fraud, identity theft, and embezzlement, according to ACFE. Cyberattacks on electronic financial services cost billions annually. In 2020, the FBI's Internet Crime Complaint Center (IC3) received over 790,000 cybercrime complaints, resulting in damages over $4.2 billion. Financial fraud has far-reaching effects beyond direct victims and enterprises. Financial fraud has a significant impact on the global economy. According to the IMF, financial crimes, including money laundering, corruption, and tax evasion, account for 2 %–5 % of global GDP, or approximately trillions of dollars. Safeguarding financial transactions and mitigating the impact of cybercrime on individuals, organizations, and the global economy requires robust fraud detection systems. Fraud detection systems have many issues. First, as digital payments and electronic services proliferate, traditional rule-based fraud detection systems struggle to keep up with thieves' innovations. These systems' predetermined criteria and thresholds make them less responsive to sophisticated fraud practices. It is increasingly challenging to verify every financial transaction manually for fraud as volumes and complexities rise. More efficient and accurate automated fraud detection technologies are needed (Federal Trade Commission, 2023).

At its core, a financial fraud detection system is an application of anomaly detection (Hernandez Aros et al., 2024). These systems are designed to monitor vast streams of transactional data and identify activities that deviate from established patterns of legitimate behavior. By creating a baseline of normal user activity, the system can flag outliers that may indicate fraudulent actions. This process is critical because fraudulent transactions often exhibit characteristics that distinguish them from a user's typical financial conduct. The goal is to detect and prevent illicit activities in real-time without inconveniencing legitimate users, a balance that presents a significant technological challenge (van Daalen, 2023).

Financial fraud causes global problems and hinders financial sector growth. A heavily skewed dataset makes it hard to detect fraud, even though the fraction of non-fraud firms is high. Thus, clever systems for detecting fraud in financial statements exist. Most approaches focus only on quantitative financial statement ratios. To evade detection, thieves employ clever evasion tactics or blend fraudulent and lawful transactions. To prevent disrupting legitimate transactions, fraud detection systems that can recognize complex fraudulent behaviors while minimizing false positives are in demand. Fraud detection methods identify unusual transactions that violate electronic services. Improvements in operational risk in financial fraud detection systems are crucial due to the significant rise in fraud and financial crimes caused by digital payments (Alfaiz & Fati, 2022). Scientists focus on antifraud; thus, they created risk assessment and detection technologies.

The growing popularity of internet banking and the diversification of payment methods have led to more sophisticated banking fraud, particularly money laundering, which harms economies at multiple levels. Old rule-based systems struggle to adapt to evolving fraud practices, resulting in high false-positive rates. Deep learning models like LLMs examine complex data to detect transaction and spending irregularities, detect fraud [Paramesha, Rane, and Rane (2024), Xu, J. (2024)]. Managing large multinational networks, real-time detection, and integrating diverse data sources, such as KYC profiles, are significant challenges. Innovative, scalable solutions are needed to address financial fraud [Usman et al. (2023)].

The massive amounts of data created by current digital payment systems may potentially challenge traditional fraud detection approaches. To better detect fraud, new technologies such as machine learning must be utilized to evaluate large datasets and derive valuable insights. Machine learning (ML) can facilitate the creation of new products and services. Researching its efficacy in several areas has advanced it. ML techniques must be cross-domain studied to determine generalizability and transferability. Identifying ML algorithm shortcomings and strengths speeds up development with less effort. Highly accurate ML algorithms can detect financial fraud in healthcare. We need cross-domain analysis to design and implement ML algorithms (Khetani et al., 2023). Swindles, fraud, and extortion threaten financial services, businesses, and governments. They threaten people, organizations, and governments (Jessica et al., 2023). Audit and Risk Committees, as well as Senior Management, must identify and implement effective measures to prevent ongoing fraud (Rehman & Hashim, 2020). Frauds pose a significant operational risk for enterprises, particularly financial firms (Mashrur et al., 2020). To prevent fraud swiftly and effectively, financial fraud detection systems are becoming invaluable.

With this backdrop, the current research objectives are

1. Improve the operational risk framework to enhance the financial fraud detection system's predictive performance.
2. Develop Machine Learning gadget (TabNet) to detect suspicious transactions in financial electronic services using a big dataset in India.
3. Improve model correctness through feature engineering.

The manuscript has the following unique features:

- **Relevance**: The topic addresses a critical and timely issue in financial fraud detection with a strong practical application.
- **Novelty**: The use of TabNet for fraud detection is still emerging, and your empirical comparison across DNN, GRU, LSTM, CNN1D, and TabNet enhances the manuscript's contribution.
- **Dataset Size and Preprocessing**: Balancing a large dataset using SMOTE enhances credibility and reproducibility.
- **Performance Metrics**: Multiple metrics, including ROC-AUC and confusion matrices, are presented clearly.
- **Visualization**: Extensive use of heatmaps, radar plots, and ROC curves increases interpretability.

Section 2 of this research examines the theoretical framework. Section 3 discusses research literature on financial fraud detection systems. We will explain our methods in Section 4. In Section 5, we present the results. Section 6 concludes this study with potential next directions.

## 2. Theoretical framework of deep neural networks (DNN) for fraud detection

### 2.1. Deep neural networks (DNNs)

Deep Neural Networks (DNNs) are hierarchical models that aim to learn complex, non-linear relationships in high-dimensional data. Based on how the human brain is organized and works, DNNs have several layers of connected nodes (neurons), with each layer doing a simple calculation followed by a more complex one. These layered architectures are well-suited for modeling intricate patterns in data, particularly where relationships among features are implicit and difficult to manually engineer**.** In financial fraud detection, transactions often involve subtle and complex interactions between features such as transaction time, frequency, device type, account balance, and merchant category. This ability to automatically learn hierarchical features makes DNNs particularly powerful for identifying novel and sophisticated fraud schemes that do not conform to predefined rules. Traditional rule-based or even shallow machine learning models may not adequately capture these intricate patterns. DNNs can effectively understand complex relationships between features and hidden patterns that are important for identifying changing fraudulent behavior.

### 2.2. Structural components of DNNs in the fraud detection context

#### 2.2.1. Input layer

The input layer receives feature vectors from the dataset, typically normalized or standardized, for processing. In the context of bank transactions, input features may include both numerical (such as transaction amount and account balance) and categorical variables (such as device type and transaction type).

#### 2.2.2. Hidden layers

These are the core computational units. Each hidden layer performs the following**:**

A hierarchical model to learn complicated, non-linear relationships in high-dimensional data is a Deep Neural Network (DNN). Deep neural networks (DNNs) mimic the architecture and operation of the human brain by using hundreds of interconnected layers of nodes (neurons) that execute a series of linear transformations and non-linear activations. When dealing with complex patterns in data, such as those with implicit and difficult-to-engineer feature interactions, layered architectures shine. Time, frequency, device kind, account balance, and merchant category are some of the features that interact subtly yet complexly with one another in financial fraud detection. Such complex patterns may be beyond the capabilities of conventional rule-based or even basic machine learning models. For detecting evolving fraudulent activity, DNNs provide the opportunity to model latent representations and high-order feature interactions.

##### 2.2.2.1. DNN structure of interest in fraud detection

*2.2.2.1.1. Layer 2.1: input.*

- Linear Transformation**: Z(l) = W(l)X(l−1) + b(l)Z^{(l)} = W^{(l)} X^{(l-1)} + b^{(l)}Z(l) = W(l)X(l−1) + b(l)**
- Non-linear Activation**: A(l) = f(Z(l))A^{(l)} = f(Z^{(l)})A(l) = f(Z(l)),** where f could be ReLU, sigmoid, or tanh.

The input layer receives normalized or standardized dataset feature vectors for processing. Numbers (such as the transaction amount or current account balance) and categories (such as the type of device used or the nature of the transaction) are both possible input features in a banking environment.

*2.2.2.2. Encapsulated (hidden) layers.* Below is a list of what each hidden layer does:

The model can learn hierarchical feature representations with deeper levels. Simple interactions (such as big transactions) may be learned by the first layer, whereas more complex combinations (such as minor transactions at high-risk hours from flagged devices) may be captured by subsequent layers.

#### 2.2.3. Output layer

The output layer usually employs a sigmoid function for binary classification.

P(Fraud) = 1+e−Z(L)P(\text{Fraud}) = \frac{1}{1 + e^{-Z^{(L)}}}P(Fraud) = 1+e−Z(L)1

The model predicts the probability of a transaction being fraudulent.

#### 2.2.4. Learning mechanism and loss function

The DNN learns by minimizing a loss function using gradient descent-based optimization.

- **Loss Function:** Binary cross-entropy is used for fraud detection.

L = −1N∑i = 1Nyilog-(y^i)+(1−yi)log-(1−y^i)\mathcal{L} = - \frac{1}{N} \sum_{i = 1}^{N} y_i \log(\hat{y}_i) + (1 - y_i) \log(1 - \hat{y}_i)L = −N1i = 1∑Nyilog(y^i)+(1−yi)log(1−y^i) where yiy_iyi is the true label and y^i\hat{y}_iy^i is the predicted probability.

- **Backpropagation:** In backpropagation, the network parameters are updated by computing and propagating the gradients of the loss function with respect to the weights.

### *2.3. DNN's advantages in identifying fraud*

- Using DNNs, complicated transactional data can be represented meaningfully without the need for costly feature engineering, thanks to automatic feature extraction.
- DNNs are scalable because they can efficiently oversee enormous volumes of data with the help of GPU acceleration.
- Capturing prevalent fraudulent behavior's non-linear connections between features through non-linear modeling.
- Adaptability to Evolving Threats: DNNs can learn from new data to identify previously unseen fraud patterns.
- Incremental Learning: DNNs can be enhanced for online learning with the right architectural tweaks, which is useful for detecting fraud in real-time.

### *2.4. Obstacles and constraints in detecting fraud using DNNs*

- Model bias toward the majority class is often the result of imbalanced data, as fraudulent transactions are rare (1–2 %). Our research proved this to be true; by default, the DNN model classified all occurrences as "fraud," resulting in a collapsed classifier with an ROC-AUC of 0.5.
- In regulatory contexts where explainability is required, DNNs are not applicable because of their lack of interpretability, which makes them operate like black boxes.
- Overfitting: Failure to regularize DNNs with dropout, batch normalization, or early halting makes them susceptible to overfitting on datasets that are noisy or imbalanced.
- Architecture that is not dynamic: regular DNNs cannot adapt to innovative ideas (such as evolving fraud patterns) without retraining, which is a significant drawback of their design.

### *2.5. Progress towards adaptive models and other options*

Modern solutions, such as Graph Neural Networks (GNNs), Autoencoder-LSTM, and TabNet, are being utilized to address these difficulties. Compared to DNNs, TabNet did far better in our study because:

- The model learned to use sparse attention to adapt to structured tabular data.
- Masks of features are used to provide interpretability.
- Overfitting can be prevented with the help of built-in regularization.

### *2.6. Theoretical justification of DNN failed in the present investigation*

One possible cause of the model's class discrimination issues is:

- Gradients Over Time: Inadequate weight initialization or an excessive number of layers can cause gradients to disappear during backpropagation, leading to subpar training.
- Inadequate Regularization and Initialization: DNNs fail to generalize because they memorize data in the absence of batch or dropout normalization.
- If class weighting or focus loss is not used, DNNs can end up learning too much from the class that doesn't have fraud, which causes an imbalance in the data learning.

### *2.7. Conclusion*

Because of their ability to represent non-linear, high-dimensional patterns, DNNs provide a theoretically robust foundation for fraud detection. Our research revealed that there are limits to these methods that must be carefully considered when putting them into practice, including data balancing, architectural choices, and interpretability. Our transition from DNN to TabNet is both theoretically and methodologically sound, in line with current tendencies in explainable AI and adaptive fraud detection systems.

## 3. Literature review

The Fintech Council of India reports that, as of 2024, over 338 lending Fintech start-ups in India are transforming the credit landscape by addressing the unmet credit needs of both rural and urban populations through innovative, data-driven lending practices. One of fintech's most significant contributions to credit risk management is the ability to utilize non-traditional data. Businesses are increasingly utilizing advanced data analysis to evaluate financial backgrounds, spending habits, and even online activity. This allows for a more complete picture of credit risk, fostering a more inclusive assessment, especially for individuals without established credit records. Digital lenders can construct strong credit profiles by examining diverse datasets, including alternative data from financial technology applications and account aggregation services. These methods offer a thorough understanding of a borrower's repayment capacity, lessening dependence on conventional credit scoring. Cloud computing is vital for fintech's efficient handling of large data volumes. Cloud-based analytical platforms empower financial institutions to process and evaluate lending risks instantaneously. A notable illustration of how fintech streamlines credit assessment is ICRA Analytics' credit risk management tool, IRS Lite. By providing organizations access to risk models across various lending categories, these solutions deliver scalability, security, and adaptability—essential components for risk management in the digital lending landscape.

Fraud Detection and Security: As fintechs broaden their scope, maintaining a competitive edge necessitates the implementation of robust fraud detection mechanisms. Financial transactions must be continuously monitored to detect anomalies and prevent fraudulent activities. Early warning systems and AI-powered solutions, such as CRisMac EWS (Early Warning System), provide fintechs with the tools to identify potential fraud and mitigate risks effectively. By automating fraud detection and reducing false positives, these systems help fintechs maintain the integrity of their lending operations while enhancing customer trust. Despite the tremendous opportunities, fintech companies face several challenges as they navigate the evolving financial landscape. Data security remains a critical concern, as fintechs oversee sensitive customer information without the physical security measures found in traditional banking. Regulatory compliance is another significant challenge, as fintechs must adhere to strict legal frameworks while remaining agile in a competitive market. Additionally, fintechs must continue refining their strategies for managing credit risk to minimize defaults while expanding their lending portfolios. Integrating AI and machine learning with big data presents technical challenges, requiring fintechs to invest in sophisticated systems and ongoing maintenance. Fintechs have been evolving alternative lending models to offer tailored products catering to specific geographies, customer segments, and industries. With faster underwriting processes and reduced loan processing times, fintechs are reshaping lending dynamics for small businesses and start-ups. To thrive in this dynamic environment, fintech companies must continue to adopt a multi-faceted approach to credit risk management. Developing data-driven credit policies that consider both traditional and non-traditional parameters will be key to making informed lending decisions (Durgesh Jaiswal (2025)).

The ability to identify suspicious transactions is one of banks' biggest concerns. Financial institutions utilize fraud detection systems to categorize transactions based on suspicion (Kannan & Srinath, 2018).

### 3.1. *Operational risk management at banks*

According to the Basel Committee on Banking Supervision (BCBS), operational risk is the risk of financial loss due to failures in internal processes, people, and systems, or external events. It is crucial to banking risk management (AL-kiyumi et al., 2021). International studies have shown that operational risk management techniques impact the financial performance of commercial banks. These findings show that understanding operational risk, including risk management in company strategy, and investing in risk management software are essential priorities. Operational risk significantly affects the performance of commercial banks (Fadun et al., 2020). The author examined how market, credit, liquidity, and operational risk affect commercial bank finances. According to the study, credit, liquidity, and operational risks negatively impact the financial performance of Kenyan commercial banks. Operational risk was most substantial (Ondigo, 2019). Additionally, Indonesian research has found that operational and liquidity risks negatively impact the financial performance of stock exchange-listed banks (Purnama et al., 2021).

**Table 1**
Summary of additional literature review.

| Study | Approach & Algorithms | Focus & contributions | Dataset used | Evaluation metrics (Accuracy %) | Findings |
|---|---|---|---|---|---|
| Esenogho et al. (2022) | Machine Learning Algorithms | Compared ML algorithms for fraud detection in blockchain | Blockchain Transaction Data | Accuracy: 90 % | Identified suitable ML algorithms for blockchain fraud detection |
| Zhang et al., 2021 | Feature Engineering Strategies | Explored feature engineering strategies for fraud detection | Credit Card Transaction Data | Accuracy: 89 % | Identified effective feature engineering strategies for fraud detection |
| Eswar (2021) | Fuzzy Rough Nearest Neighbor & Logistic Regression | Utilized fuzzy logic for fraud detection | Credit Card Transaction Data | Accuracy: 91 % | Demonstrated the effectiveness of fuzzy logic in fraud detection |
| Hussein et al. (2021) | Web Service-Based Fraud Detection | Proposed a web service-based approach for fraud detection | Web Service Data | Accuracy: 90 % | Demonstrated the feasibility of web service-based fraud detection |
| Nandi et al. (2022) | Network-Based Extensions | Proposed network-based extensions for fraud detection | Credit Card Transaction Data | Accuracy: 88 % | Developed a novel approach for automated fraud detection |

### 3.2. Financial fraud detection system

While earlier methods struggle to adapt to changing fraud trends, Alatawi (2025) used a carefully selected set of innovative ML algorithms to address the complexities. These models include neural networks, decision trees, SVMs, random forests, and clustering. Big data processing and cloud computing enable real-time fraud detection by efficiently evaluating enormous volumes of transactional data. Using feature engineering and anomaly detection, the model can identify fraudulent transactions. Comprehensive trials on a broad and diverse set of real and simulated credit card transactions, including both valid and fraudulent transactions, confirm that this technique is effective. The results demonstrate state-of-the-art fraud detection, achieving higher precision and recall than classic approaches. The study discusses ML models that enhance the management and prevention of fraud risk. The findings of this study help banks, government organizations, and regulators fight credit card fraud's harmful consequences on customers, companies, and the economy. Using a novel ML-driven strategy, this study solves the IoT ecosystem fraud problem and lays the path for future advances in this vital area.

Alarfaj et al. (2022) used innovative ML and DL algorithms to detect fraud, whereas Alharbi et al. (2022) developed a DL-based text2IMG technique. Bagga et al. (2020) and Baker, Mahmood, and Shaker (2022) studied ensemble learning. Auto loan fraud detection (Randhawa et al., 2018), shill bidding fraud detection (Madhurya et al., 2022), and fuzzy rough nearest neighbor credit card fraud detection were studied. Table 1 Summarises additional recent studies.

Alatawi (2025) shows that Random Forest and Gradient Boosting Machine increase real-time fraud detection. Financial organizations can utilize these models to minimize false positives, enhance security, and respond promptly to suspicious activity, thereby limiting financial losses and preserving consumer trust. Integrating IoT data into these systems allows constant monitoring and updating, making them adaptive to new threats.

According to Kurshan et al. (2020), graph-based solutions for digital transaction data fraud detection can be complicated due to the size, speed, and nature of financial crime detection applications. Adversarial techniques will be difficult; hence, graph-based solutions should be improved. As internet use and online businesses rise, financial frauds increase, hurting the economy. Some machine learning and data mining methods are being employed to address this issue; however, calculation speed, processing vast amounts of data, and discovering unidentified attack patterns require refinement.

Alghofaili et al., 2020 developed a deep learning-based financial fraud detection system using Long Short-Term Memory (LSTM). The authors evaluated the LSTM on a credit card fraud dataset and compared it to the auto-encoder model and other machine learning methods. The LSTM achieved 99.95 % accuracy in under a minute, demonstrating its exceptional performance.

Statistical and machine learning methods for detecting financial crimes have been extensively studied (Leo et al., 2019). Machine learning evaluates historical data to improve performance or predictability (Nathani & Singh, 2021). Operational domains use machine learning to identify and prevent risks. ML focuses on fraud detection and suspicious transactions in operational risk, not cybersecurity (Khrestina et al., 2017).

Alfaiz and Fati (2022) used ML to block all malicious credit card transactions. For the banking transaction network, Khosravi et al. (2023) developed fraud prediction models. To do this, they utilized various supervised ML approaches. They used 46,316 client transactions to assess these methods. 25 transaction network functionalities were also incorporated. Feature with a Pearson correlation larger than 0.8 was removed. The models' accuracy, recall, precision, and F1 score were assessed. According to the study, ML models can detect organized banking fraud. The above experiments offer a highly accurate method for detecting credit card fraud that adapts to transaction type and consumer characteristics.

Automation has led to the development of diverse tools for fraud detection, as briefly discussed herein. By eliminating manual intervention and saving money, automation accelerates fraud detection [Aslam et al. (2022)]. Subsampling selects representative datasets, reducing computational costs and retaining feature correlations [Siddamsetti & Srivenkatesh (2024)]. VFDT rapidly oversees real-time data streams and performs well with blockchain technology, ensuring secure and scalable updates [Dhieb et al. (2020)].

*Blockchain smart contracts allow for safe and shared data without a central authority, helping to quickly spot fraud and adjust models as needed.* Automatic parameter adjustment [60] and resampling hybrid approaches [Khashan (2024); Mustafa Abdul Salam et al. (2024)] enhance detection accuracy, whereas Stochastic Gradient Descent (SGD) incrementally optimizes models [Hemant Palivela et al. (2024)]. Distilling knowledge from sophisticated models to lightweight ones enables real-time detection. Preprocessing and model training are included in fraud detection pipelines. Classifiers like the Random Forest Quantile Classifier optimize sensitivity and specificity for imbalanced data [CARRACEDO et al. (2024)], making stacking ensembles more resilient [Zhang et al. (2021)]. These developments enable scalable, adaptive, and accurate financial fraud detection systems that can manage complicated fraud.

*Convolutional Neural Networks (CNNs) find unusual spending habits or risks linked to specific merchants by looking at complex data, like time-series patterns and transaction maps. It automatically picks out important details and simplifies the data, so it needs less preparation before training.* It automatically captures critical features and reduces feature dimension; thus, training requires little data preprocessing [Alarfaj et al. (2022)]. When combined with standard models (SVM, KNN, NB, and DT), CNN deep features enhance fraud detection [Ming et al. (2024)].

*Recurrent Neural Networks (RNNs):* RNNs can analyze transaction histories because they preserve context from earlier inputs. However, gradient vanishing and ballooning gradient concerns limit their capacity. GRU and LSTM are specialized RNNs that reduce gradient vanishing and explosive gradients [Almazroi & Ayub (2023)].

*Multi-Layer Perceptrons (MLPs)*: MLPs model structured data interactions using interconnected layers. MLP is competitive for fraud detection in amount-based profiling and non-linear data problems [Can et al. (2020)].

*Transformers*: Transformers process the entire input sequence at once, unlike CNNs and RNNs. Transformers can identify patterns in transaction data, user actions, and network connections for fraud detection by using their self-attention mechanism and feed-forward networks to grasp complex relationships and highlight important features from sequential data.

*Natural Language Processing (NLP):* By evaluating unstructured textual data such as financial reports [Zhang et al. (2022)], tax compliance and financial regulation [Bajpai (2024)], and claims narratives [Fursov et al. (2022)], NLP improves fraud detection. Readability measurements, sentiment analysis, and feature extraction (Bag-of-Words, TF-IDF, and word embeddings) are important. These features enhance accuracy and recall in machine learning models with classical fraud indicators [Zhang et al. (2022)].

### *3.2.1. Variational autoencoders (VAEs) and generative adversarial networks (GANs)*

Popular deep learning generative models include GANs and VAEs [Jiang et al. (2023)]. GANs provide realistic samples but are challenging to train because of adversarial dynamics and a non-interpretable latent space. With a well-structured latent space, VAEs are superior to GANs for representation learning and probabilistic modeling, although they produce lower-quality samples. Combining VAEs and GANs can overcome their limitations in managing imbalanced data for fraud detection [Ding et al. (2023)].

*Graph Neural Networks (GNNs)*: The graphs can model complicated interactions and relationships between entities (e.g., transactions, reviews), making GNNs effective fraud detectors [Nouhaila Innan et al. (2024)]. Analyzing these linkages can reveal fraud tendencies and anomalies [Shehnepoor et al. (2022)]. One type of GNN uses graph convolution. A study found that GCNs outperform standard models like LR, RF, and GBMs in detecting fraudulent transactions [Usman et al. (2023)].

*Deep Belief Networks (DBNs)*: Due to their feature extraction and classification capabilities, DBNs are a key method for deep learning-based fraud detection. They are less effective than CNNs, which are more accurate and scalable in fraud detection [Alarfaj et al. (2022)].

*Machine Learning (ML) Models:* Traditional ML methods are used to detect financial fraud [Cherkaoui et al. (2024)]. A simple and interpretable binary classification method is Logistic Regression (LR). While LinearSVC efficiently processes linear data, KNN finds anomalies in proximity but lacks scalability for large datasets. Random Forests and Gradient Boosting (XGBoost, LightGBM) ensemble approaches increase accuracy by mixing models. Random Forests prevent overfitting, while Gradient Boosting balances datasets. AdaBoost iteratively improves poor classifiers for fraud. High-dimensional SVMs recognize outliers well. Although they can be prone to overfitting, decision trees are interpretable and essential for ensemble models. These methods offer standards for structured data performance and enhancements to fraud detection systems, despite being less flexible than deep learning.

*Hybrid Models*: For complicated tasks like financial fraud detection, hybrid models combine algorithm strengths. ASA-GNN makes GNN frameworks better by using smart sampling and a method that groups information based on uncertainty, which helps solve the main issues regular GNNs have in finding fraud. ASA-GNN may detect complicated fraud patterns in graph-structured data by focusing on important neighbors and avoiding over-smoothing [Tian et al. (2023)].

*Reinforcement Learning with Deep Q-Network (RDQN):* Deep learning and reinforcement learning are integrated in RDQN. RDQN uses Rough Set Theory for feature reduction and reinforcement learning to process faster and more accurately, making it a scalable and effective fraud detection solution. RDQN beats SVM, ANN, DT, IFDTC4.5, SAE-GAN, and CNN-SVM-KNN [Chandana Gouri & Natarajan (2023)].

The Transformer-LOF-Random Forest model finds complex and unusual fraud patterns by using better feature extraction, local anomaly detection, and combining different learning methods. It improves upon XGBoost, LightGBM, and LSTM by addressing data imbalance, reducing false positives and negatives, and responding to new fraud methods [Miao (2024)].

ResNeXt-embedded Gated Recurrent Unit (RXT-J) combines ResNeXt's feature extraction and GRU sequential learning. RXT-J beats BERT, ANN, and logistic regression [Almazroi & Ayub (2023)].

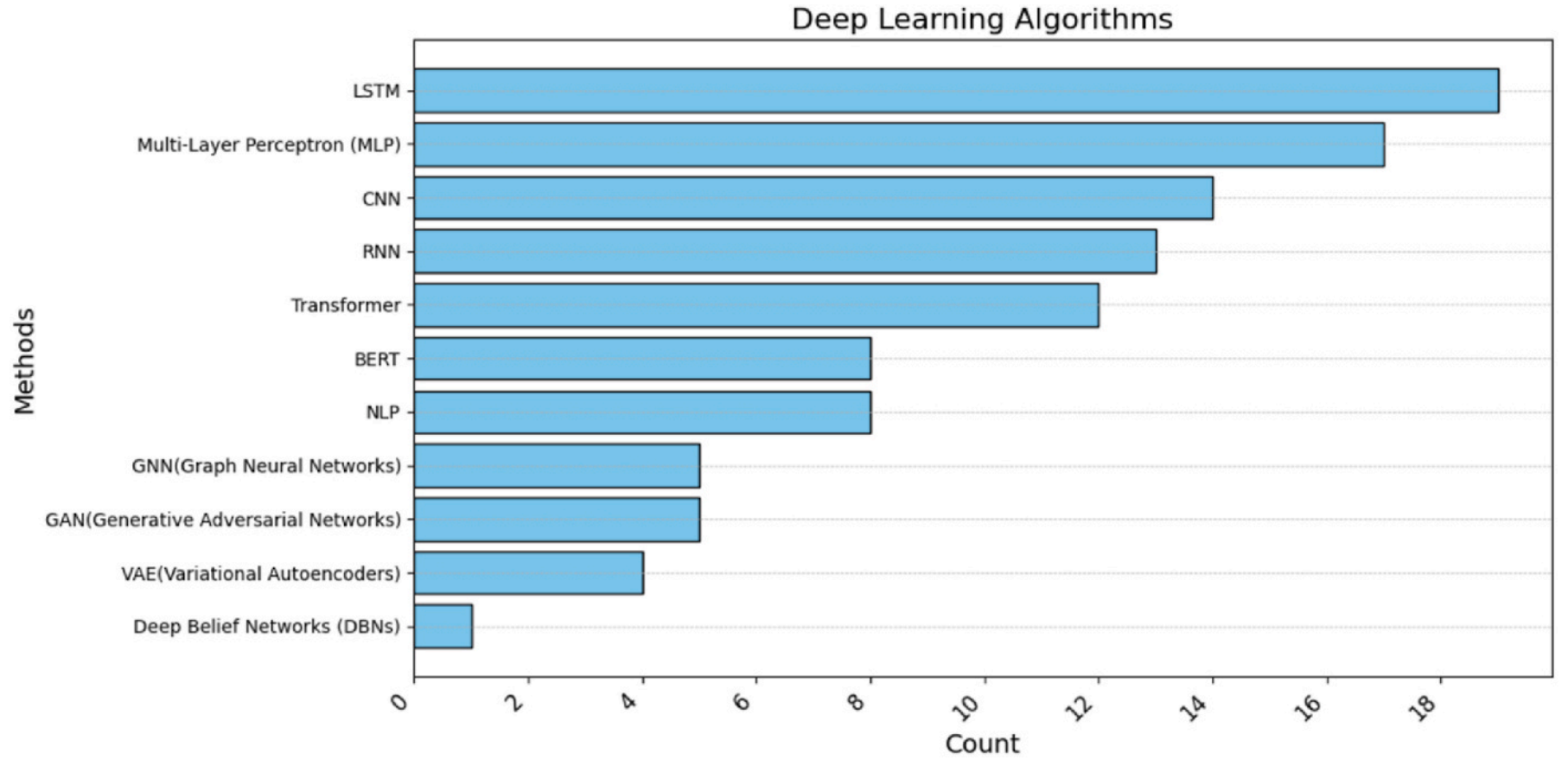

**Fig. 1.** Distribution of deep learning techniques applied to financial fraud detection.
(Source: Chen et al. (2025))

CatBoost-Deep Neural Networks integrates CatBoost with Deep Neural Networks (DNN) to enhance their combined capabilities. CatBoost excels at handling imbalanced datasets, categorical features, and complex structured data interactions [36]. DNN learns patterns in raw features and adapts to sparse data. Hybrid models outperform random forests and ensemble approaches, such as LSTM-based AdaBoost [Nguyen et al. (2022)].

Autoencoder-LSTM combines the deep learning models Autoencoder and LSTM. The autoencoder reduces dimensionality while preserving critical characteristics and reducing noise. The LSTM network identifies fraudulent and genuine transactions based on temporal connections. Together, the model outperforms classical machine learning and solo LSTM models [Zioviris et al. (2024)].

These diverse and increasingly complex models illustrate a clear trend in the literature: a technological arms race between financial institutions and malicious actors (Walter (2024)). As fraudsters develop more sophisticated methods, this dynamic becomes particularly relevant in technologically advanced financial markets like Austria, where banks and fintech companies must innovate to combat cross-border fraud and adhere to stringent EU regulations. The move from regular machine learning models to more advanced deep learning systems like GNNs, Transformers, and special hybrid systems is a direct response to the need to look at not just single transactions but also the networks and messy situations where fraud happens.

Fig. 1 shows the entire deployment pattern of deep learning models. The most popular are LSTM, MLP, CNN, and RNN. For textual datasets, NLP approaches like the BERT model are moderately used. Additionally, researchers have used GNNs, GANs, and VAEs to detect fraud. Since fraud detection datasets are imbalanced and GANs are modestly used, their potential for modeling relationships and creating synthetic datasets should spur further research.

Financial sectors frequently distribute deep learning algorithms to identify financial fraud, as shown in Fig. 2. Credit card and banking industries employ numerous deep learning methods. MLP.

LSTM and CNN appear to be the most popular. Due to the sequential and graph-structured nature of blockchain data, LSTM and GNN are widely used to represent transaction relationships. Tax and Money Laundering use deep learning to a limited extent. For structured data fraud detection, classical ML approaches like Random Forest and SVM are essential. Analyzing sequential and text-based fraud data with LSTM, Transformers, and BERT is growing. GNNs and GANs are getting popular in blockchain and synthetic data fraud. CNN is for credit card fraud, and GNN is for blockchain**.**

Sadgali et al. (2020) introduced a framework for detecting credit card fraud. Their research included human behavior, uneven data, and odd transactions that typical approaches could not account for. They used several detection algorithms to improve accuracy and had four components: data storage, decision-making, behavior analysis, and authentication filtering.

Table 2 Summarises the past research work that addresses these requirements.

In summary, a good fraud detection system requires the following elements (Kalbande et al., 2021):

- Accuracy: The system should detect fraudulent transactions with minimal false positives and negatives.
- Enable real-time fraud detection to avert financial losses and reduce risks.

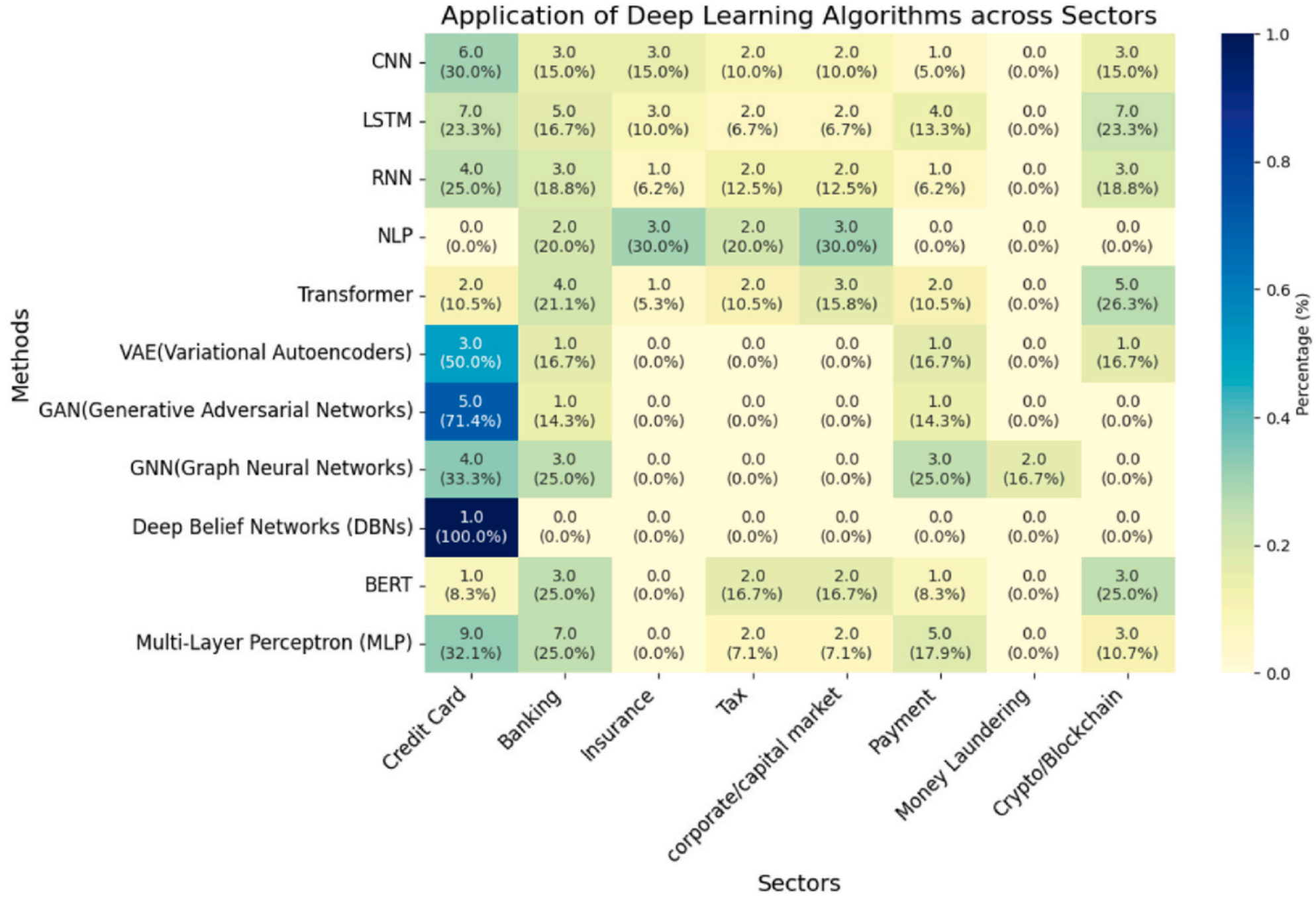

**Fig. 2.** Application of deep learning methods in financial fraud detection across sectors (Source: Chen et al. (2025)).

**Table 2**
The summary of related papers in terms of fraud detection systems requirements.

| References | Requirements for fraud detection systems | | | | | | |
|---|---|---|---|---|---|---|---|
| | Accuracy | Real-time detection | Scalability | Adaptability | Integration | Explainability | Compliance |
| Kurshan et al. (2020) | ✓ | x | ✓ | ✓ | x | ✓ | – |
| Alghofaili et al. (2020) | ✓ | x | x | ✓ | x | ✓ | – |
| Xiuguo and Shengyong (2022) | ✓ | x | x | x | ✓ | ✓ | – |
| Alfaiz and Fati (2022) | ✓ | x | x | x | x | ✓ | X |
| Alkhalili et al. (2021) | ✓ | x | ✓ | ✓ | ✓ | x | ✓ |
| Chen (2022) | ✓ | x | ✓ | ✓ | ✓ | – | – |
| Khosravi et al. (2023) | ✓ | x | ✓ | ✓ | ✓ | x | – |
| Ashfaq et al. (2022a, 2022b) | ✓ | x | x | ✓ | – | x | – |
| Amala Dhaya and Ravi (2021) | ✓ | x | x | x | – | x | – |

- Scalability: The system must manage enormous data and transaction volumes, as banking systems typically process a high volume of transactions.
- Adaptability: The system should be able to adjust to evolving fraud patterns and methods. New data should help it discover fraud patterns and upgrade its models.
- Integration: The solution should seamlessly interact with existing banking systems and processes to monitor and detect fraudulent efforts.
- The system ought to elucidate the factors that lead to the classification of a transaction as fraudulent. Use this information to understand the system's decisions and investigate fraud.
- Compliance: The system must adhere to legal and ethical norms, including anti-money laundering (AML) legislation.

### *3.3. Research gap*

The financial sector drives economic growth and stability worldwide. Financial fraud has increased due to the complexity of financial transactions and the widespread use of digital technology. Financial system fraud detection requires particular conditions. As seen in Table 1, various articles have addressed these needs. Based on earlier research, ML algorithms for fraud detection have become a key technique to combat financial crime, as demonstrated in Table 1. Understanding that numerous factors can impact the effectiveness of ML models in detecting fraud is crucial. The lack of authentic data sets, imbalanced data sets, the quantity of data sets, the caliber and inclusivity of the data set, feature selection, hyperparameter determination, and the dynamic nature of fraud are these variables. Our paper discusses how ML can detect financial system fraud and improve security. Our key research objectives stated in section 1 are implemented through the following modeling strategies:

| Our research Objectives stated in Section 1 | Modeling Strategy with bank data from India |
|---|---|
| 1. Improve the operational risk framework to enhance financial fraud detection systems prediction performance.<br>2. Improve the operational risk framework to enhance financial fraud detection systems prediction performance.<br>3. Improve model correctness through feature engineering. | • By optimizing detection rate while minimizing false positives. Models are calculated from supervised detection samples of fraudulent and legitimate transactions.<br>• By developing Machine Learning gadget to detect suspicious transactions in financial electronic services using a big dataset in India.<br>• By comparing Indian bank fraud data with other studies.<br>• Implementing requirements to enhance financial fraud detection efficiency levels |

## 4. Methodology

### *4.1. Adaptive deep learning models*

Financial organizations use risk prediction algorithms for credit scores and loan defaults to make data-driven decisions. Applying static machine learning models in changing economic situations is challenging. The model may drift as more data becomes available. Changes in distributions or patterns might affect a model's prediction performance. [1] During a recession, default rates may rise, contradicting previous data patterns. Simultaneously, deep learning models appear as opaque decision-making tools. Lack of transparency in credit risk modeling raises questions about fairness and regulatory compliance. It is crucial to have risk prediction systems that can adapt to changing facts and make transparent judgments.

Banking systems are frequently targeted by fraudsters attempting unauthorized transactions. Traditional machine learning methods often fall short in capturing complex feature interactions or explaining model decisions. Hence, the problem addressed here is the classification of transactions as fraudulent or legitimate using a deep learning model that can natively oversee tabular data, offer interpretability, and cope with class imbalance. The objective of this research is to develop an effective, interpretable, and high-performing model for detecting fraudulent bank transactions using TabNet, a deep learning model tailored for tabular data. The goal is to leverage TabNet's sparse attention and feature selection capabilities to achieve high accuracy and interpretability on a real-world, imbalanced dataset.

In non-stationary situations, catastrophic forgetting can cause machine learning models to perform poorly on older data immediately after learning new data or tasks. Neural networks optimize new patterns and overwrite existing knowledge, a well-known issue. Lifelong or continuous learning enables models to learn without losing previously acquired information. Researchers have created regularization-based and replay-based neural network continual learning approaches (AR5iv.org).

### 4.2. Data

Through specific search parameters focusing on size, completeness, and relevance to our fraud detection objectives, we identified a large Kaggle dataset of valid and fraudulent transactions. Given the dataset's imbalance in fraudulent transactions, our selection was crucial. We followed data sharing and privacy ethics after carefully reviewing the dataset documentation. Initial analyses and visualizations helped comprehend the dataset's structure, characteristics, and constraints, easing data pre-processing. Kaggle's datasets strengthened our Bank Transaction fraud detection algorithm, providing detailed features. Fraudulent online and digital banking transactions pose a significant threat to financial organizations and their customers. Unauthorized account access, identity theft, and strange transaction patterns affect customers and the economy. This research attempts real-time fraud detection and preservation to address this developing issue. Old transaction data, including account information, transaction quantities, service provider records, and timestamps, is checked for fraud. A powerful fraud detection device with minimal false positives is the goal, one that can distinguish between lawful and likely fraudulent transactions. Transaction history helps device learning algorithms recognize fraud schemes and adapt to new threats. The device notifies bank staff of suspicious transactions in real time. Fraud detection enhancements safeguard consumer assets, prevent financial losses, and preserve a trustworthy financial organization. Kaggle's data features are in Table 3. These features are essential for identifying fraudulent bank transactions and training and evaluating our machine learning models.

These feature descriptions give a clear expertise of the facts to be used for fraud detection analysis. This study's temporal frequencies were vital for real-time fraud detection. To capture transaction temporal patterns in an IoT-enabled environment where transactions occur continuously and in real time, time-frequency characteristics were chosen. Fraudulent transactions often involve multiple high-value transactions within a short period, or unexpected purchases made at unusual hours. Models can detect anomalies better by studying transaction time frequencies. When combined with geographical location or store information, the interval between transactions and frequency of purchases over a period can indicate suspicious activity. Based on financial fraud patterns, we chose time frequencies.

### 4.3. Dataset preparation and description

The dataset consists of 124,496 records (after balancing using SMOTE (Synthetic Minority oversampling technique), equally split between fraudulent and legitimate transactions. Key features include.

1. Transaction Amount—Monetary value of the transaction
2. Account Balance—Balance at the time of transaction
3. Hour, DayOfWeek, Month—Temporal metadata
4. Transaction Device, Device Type—Device-related information
5. Merchant Category, Transaction Type—Categorical descriptors
6. The target variable, Is fraud, is binary: 0 for non-fraudulent and 1 for fraudulent.

**Table 3**
Description of fraud transaction data (India).

| Sl.No | Feature | Description |
|---|---|---|
| 1 | Gender | The gender of the consumer (e.g., Male, Female, Other). |
| 2 | Age | The age of the consumer at the time of the transaction. |
| 3 | State | The nation in which the patron resides. |
| 4 | City | The metropolis wherein the client is living. |
| 5 | Bank_Branch | The consumer maintains their account at a specific financial institution branch. |
| 6 | Account_Type | The kind of account held with the aid of the customer (e.g., Savings, Checking). |
| 7 | Transaction_Date | The date on which the transaction passed off. |
| 8 | Transaction_Time | The specific time the transaction was initiated. |
| 9 | Transaction_Amount | The financial value of the transaction. |
| 10 | Merchant_ID | The transaction includes a specific identifier for the merchant involved. |
| 11 | Transaction_Type | The nature of the transaction (e.g., Withdrawal, Deposit, Transfer). |
| 12 | Merchant_Category | The class of the merchant (e.g., Retail, Online, Travel). |
| 13 | Account_Balance | The balance of the customer's account after the transaction. |
| 14 | Transaction_Device | The tool utilized by the consumer to perform the transaction (e.g., Mobile, Desktop). |
| 15 | Transaction_Location | The transaction's geographical location, such as its latitude and longitude, is crucial. |
| 16 | Device_Type | The kind of device used for the transaction (e.g., Smartphone, Laptop). |
| 17 | Fraud (Target Variable) | A binary indicator (1 or zero) indicating whether the transaction is fraudulent or now not. |
| 18 | Transaction_Currency | The currency used for the transaction (e.g., Indian Rupee in this case). |
| 19 | Transaction_Description | A brief description of the transaction (e.g., buy, switch). |

The original dataset comprises features with the class label as “Is_fraud,” but the class labels are imbalanced, with Class 0 and Class 1 counting 62248 and 3287, respectively. Henceforth, the dataset was balanced using SMOTE to balance the class labels. The balanced dataset has a class 0 and class 1 count of 62248 and 62248, respectively. Figs. 3a and 3b show the class distribution of the class label before and after applying SMOTE.

Fig. 4 shows the Transaction amount alone is not a strong differentiator of fraud.

Fig. 5 shows the transaction frequency by Hour (Fraud vs non-fraud). The key insights drawn from this are as follows:

1. Most transactions occur between 8 a.m. and 5 p.m., aligning with typical banking/business hours.
2. Highest volume appears around 10–11 a.m.
3. Suggests fraud is not restricted to odd hours; it is **well-distributed.**

Contrary to expectations, fraud is not concentrated during off-hours. Most transactions, including fraudulent ones, occur between 8 a.m. and 5 p.m., with a peak around 10–11 a.m.

Fig. 6 indicates that a sizeable number of fraud cases occurred on Wednesday and Thursday. This is highly unusual, suggesting that fraudsters may be exploiting high transaction volumes midweek to evade detection.

Fig. 7 shows that Fraudulent and non-fraudulent transactions have nearly identical account balance distributions, implying that Transaction Amount alone does not strongly indicate fraudulent behavior.

Fig. 8 presents the heatmap that shows the number of fraud transactions by hour of day and day of the week. It is a powerful visualization for uncovering temporal patterns in fraudulent behavior. The heatmap shows significantly distinct temporal hotspots, particularly high in fraud activity on weekdays during late mornings and early afternoons.

Fig. 9 shows the radar plot (also called a spider plot or web chart), showing the normalized mean values of selected features, grouped by the binary class Fraud.

The radar plot shown in Fig. 9 provides better insights into feature interactions by comparing the normalized mean values of key features between fraudulent and non-fraudulent transactions. A radar plot shows that while individual features may not differentiate fraud well, combinations such as lower transaction amounts with higher account balances and specific timing can reveal meaningful patterns. It reveals that fraudulent transactions are typically associated with higher account balances and lower transaction amounts, suggesting that fraudsters may attempt to avoid detection by making smaller withdrawals from high-value accounts. Additionally, fraud tends to occur during off-peak hours and is more concentrated on specific days of the week, whereas non-fraudulent transactions are more evenly distributed across both time and days.

### *4.4. Justification for choosing TabNet over another deep neural network (DNN)*

Consistent with the discussion in the theoretical section of this research, the foregoing EDA (Exploratory Data Analysis) lays a strong groundwork for building a context-aware, interpretable, and robust fraud detection system. It calls for a change in thinking away from traditional rule-based or black-box detection approaches toward interpretable AI models, such as TabNet, which are capable of dynamic learning from complex feature interactions. The insights not only enhance predictive accuracy but also support regulatory compliance, risk management, and operational intelligence.

Traditional deep learning architectures, such as Multi-Layer Perceptrons (MLPs), often struggle with tabular data because they do not natively exploit the **sparse and structured nature** of such datasets. In contrast, **TabNet**, a continuous deep learning architecture (Fig. 1) designed specifically for tabular data, offers several critical advantages that make it more suitable for bank fraud detection:

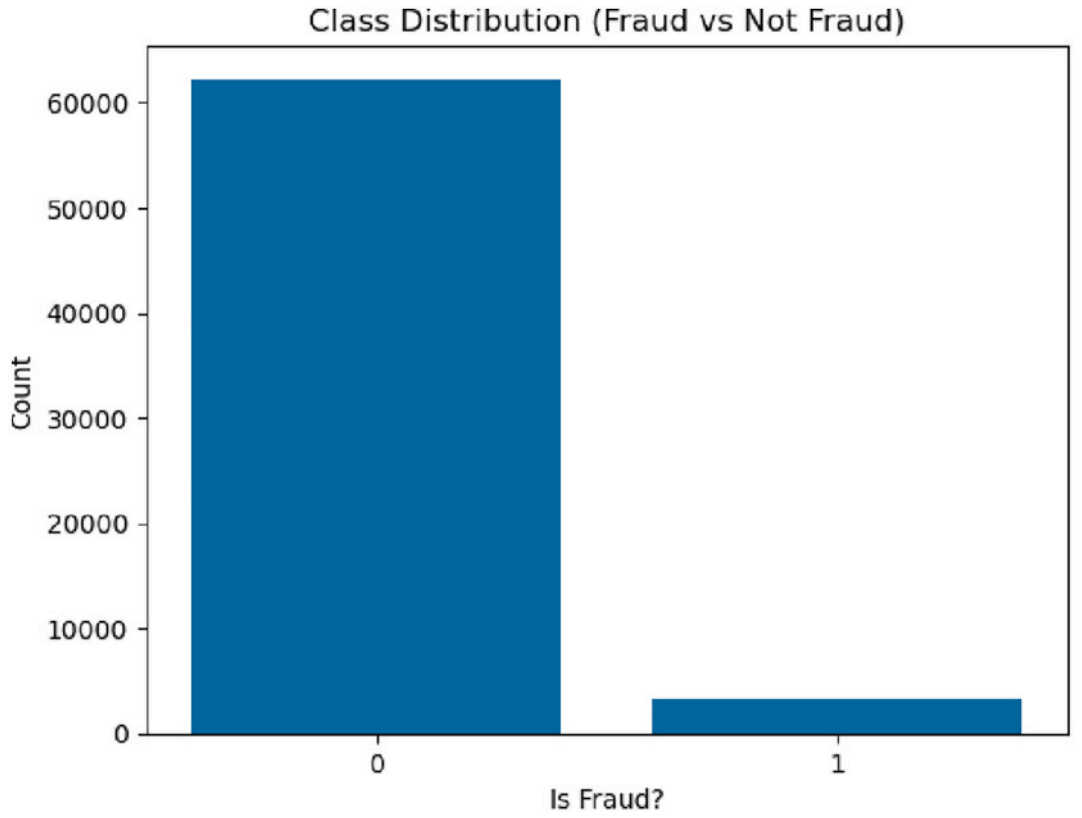

**Fig. 3a.** Class distribution of Bank Fraud dataset before SMOTE.

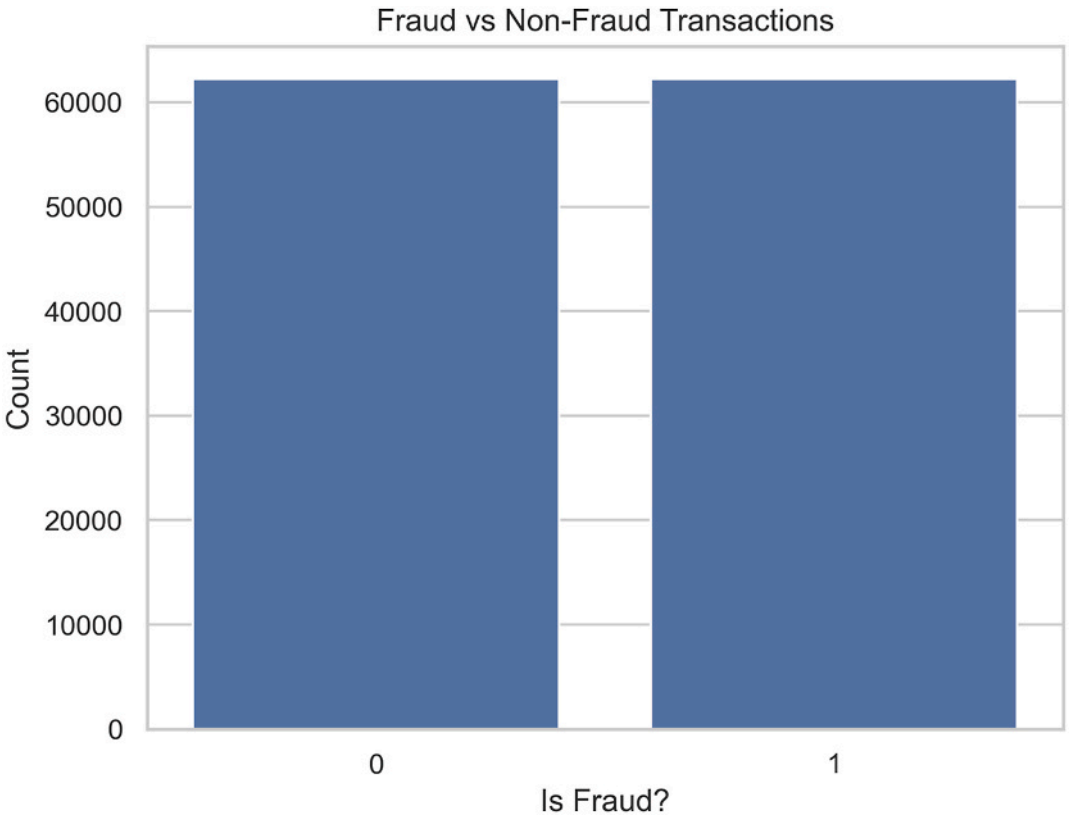

**Fig. 3b.** Class distribution of Bank Fraud dataset after SMOTE.

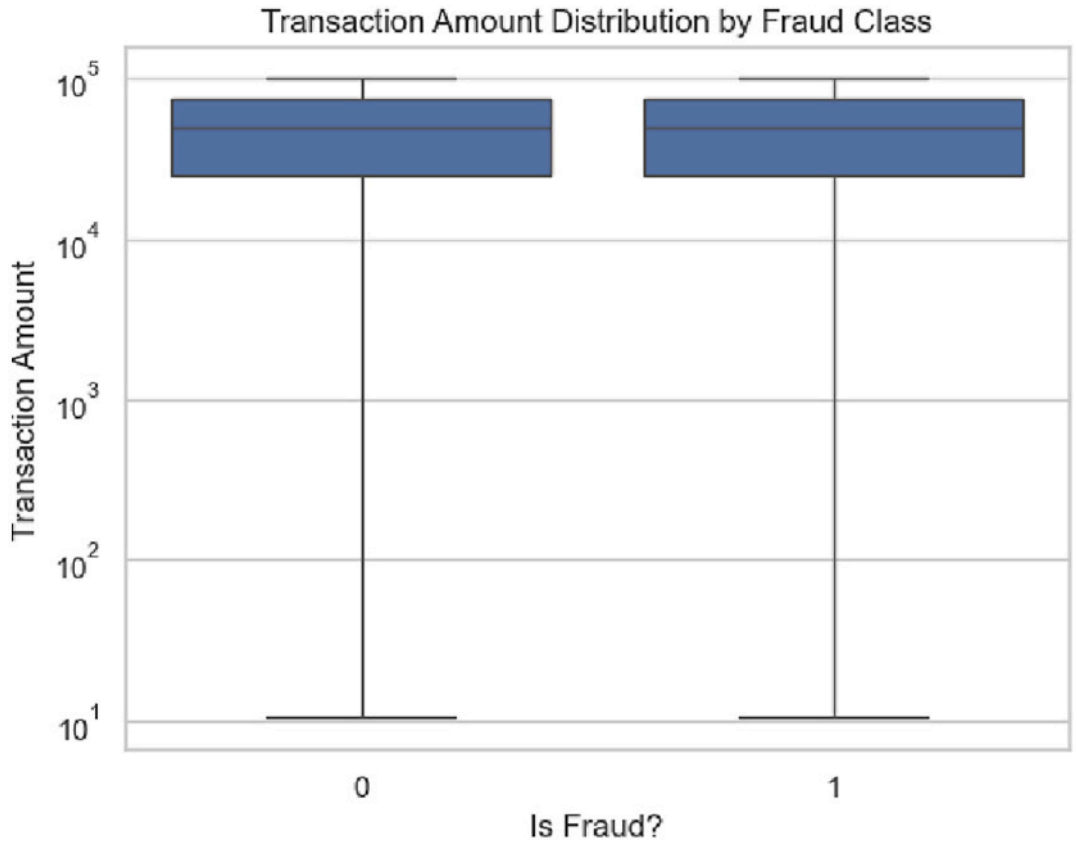

**Fig. 4.** Transaction amount distribution by Fraud Class.

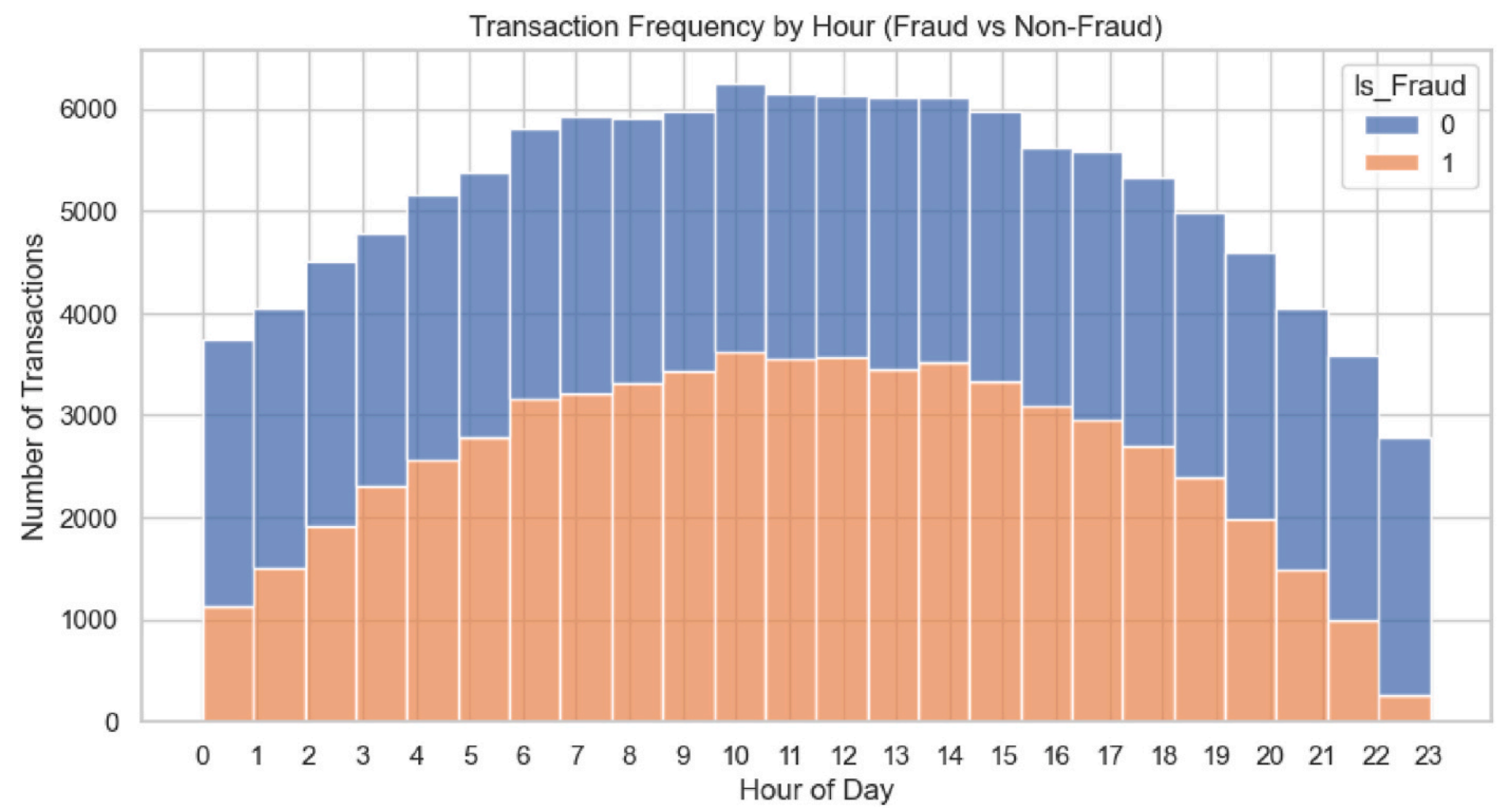

**Fig. 5.** Transaction frequency by Hour (Fraud vs non-fraud).

1. **Feature Selection via Sequential Attention**: TabNet employs a novel **sparse attention mechanism** that dynamically selects the most relevant features at each decision step. This enables the model to **focus on meaningful features** for fraud detection, thereby improving both interpretability and performance.

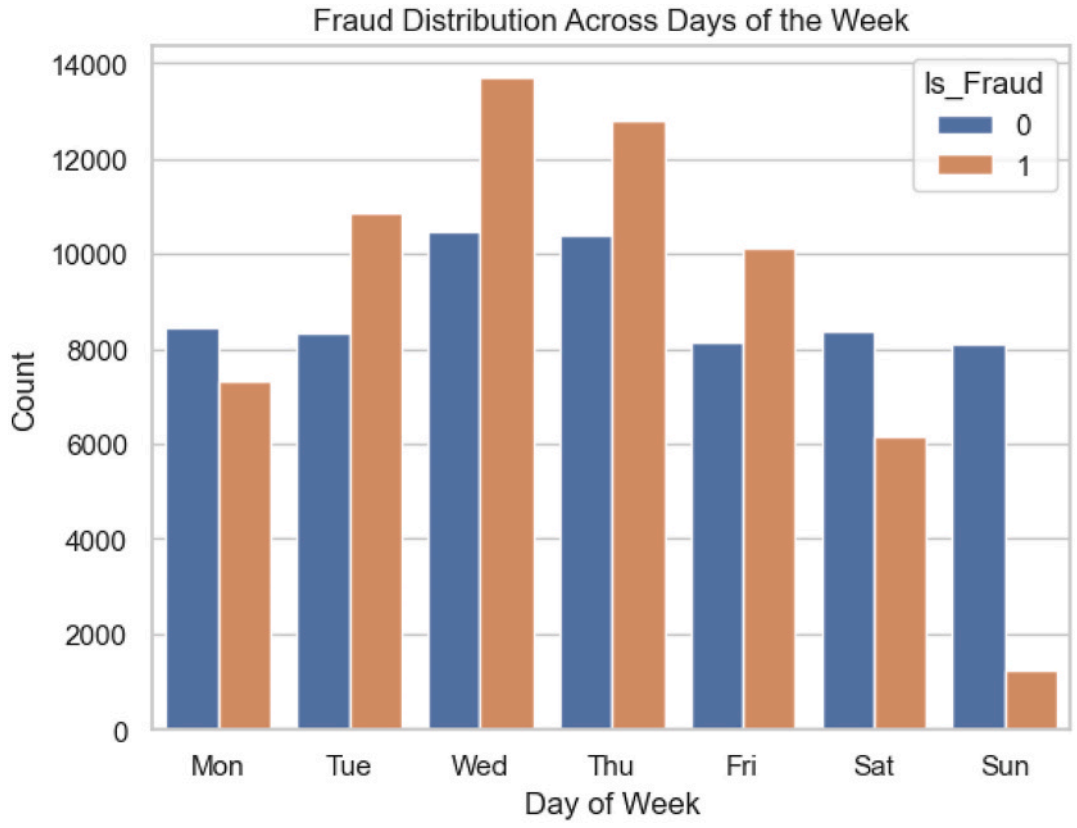

**Fig. 6.** Fraud distribution across days of the week.

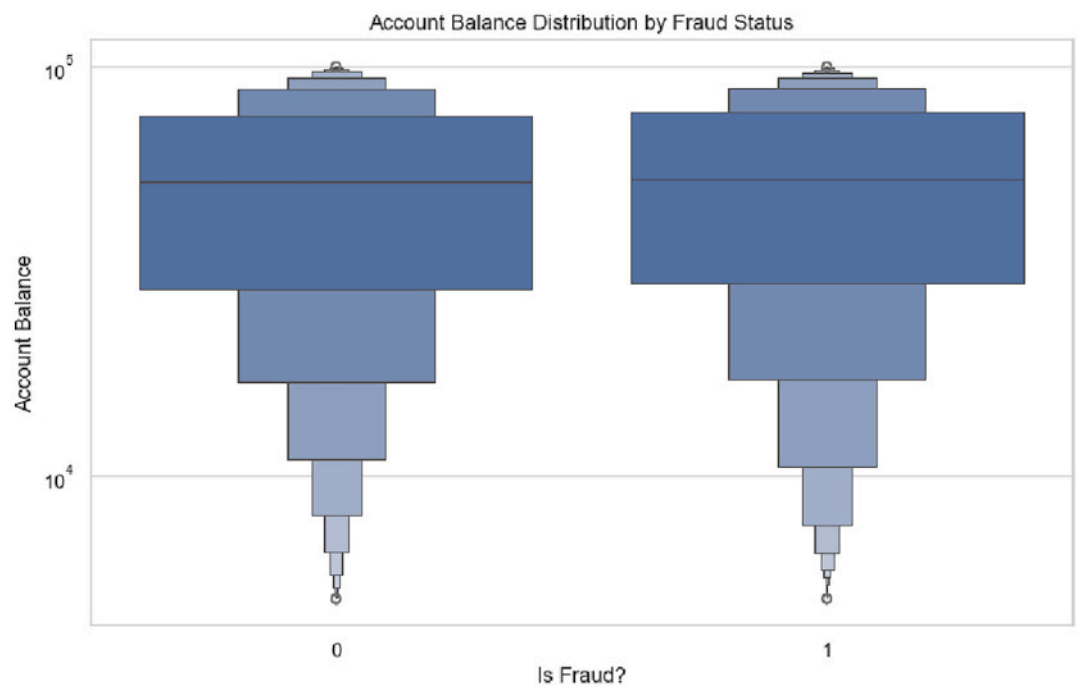

**Fig. 7.** Account balance distribution by fraud Status.

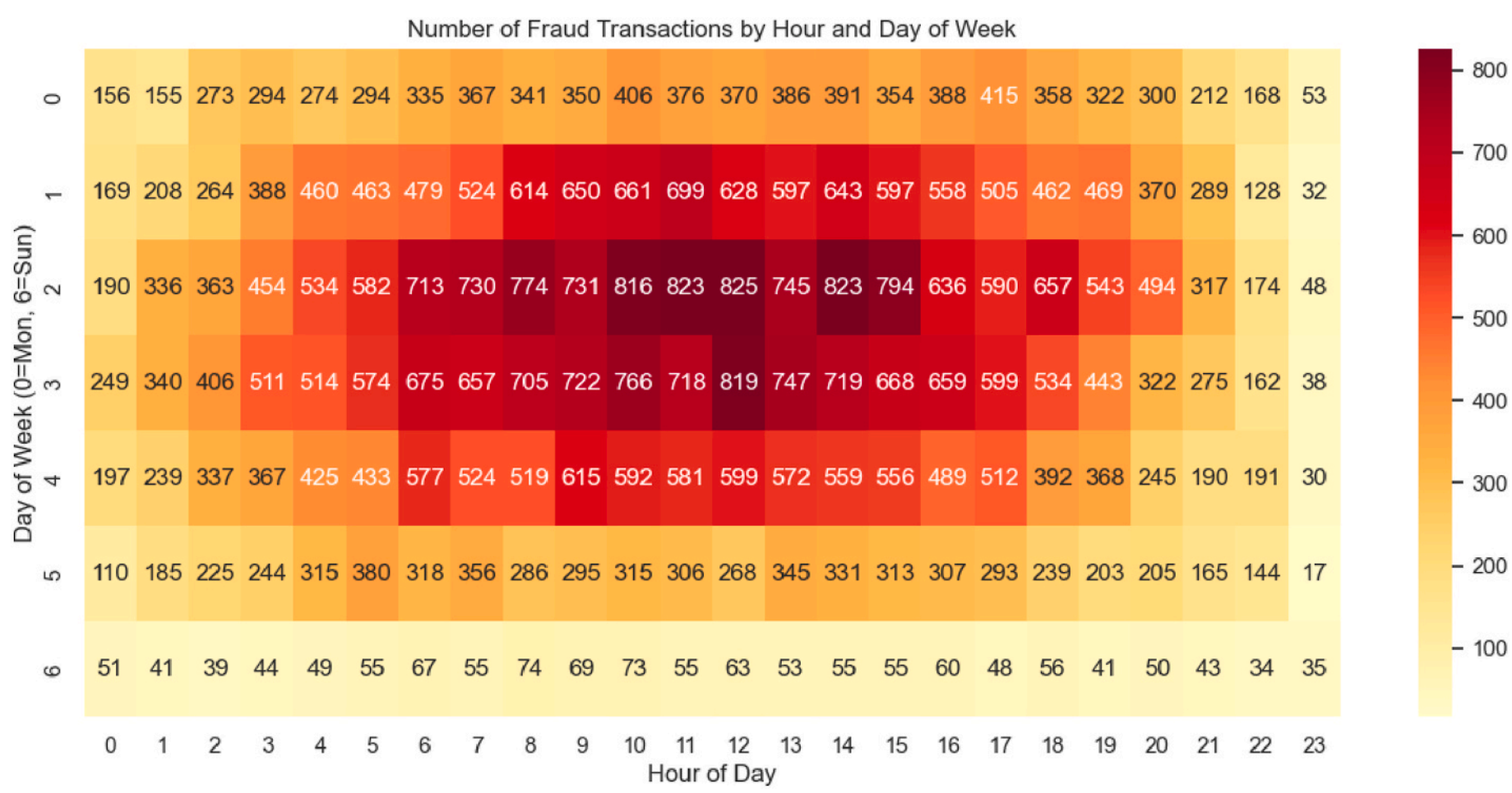

**Fig. 8.** Heatmap of the Number of fraud transactions by hour of day and day of the week.

2. **Interpretability**: One of the most significant limitations of traditional deep neural networks is their **black-box nature.** TabNet addresses this by providing **built-in interpretability**, offering insights into which features were used at each decision step. This is crucial in high-stakes domains like fraud detection, where **explainable AI** is necessary for regulatory compliance and trust.
3. **Efficient Learning on Tabular Data**: Unlike CNNs or RNNs, which excel on image and sequence data, respectively, TabNet is **optimized for structured, tabular datasets.** It efficiently models feature interactions without needing heavy feature engineering or complex architectures.

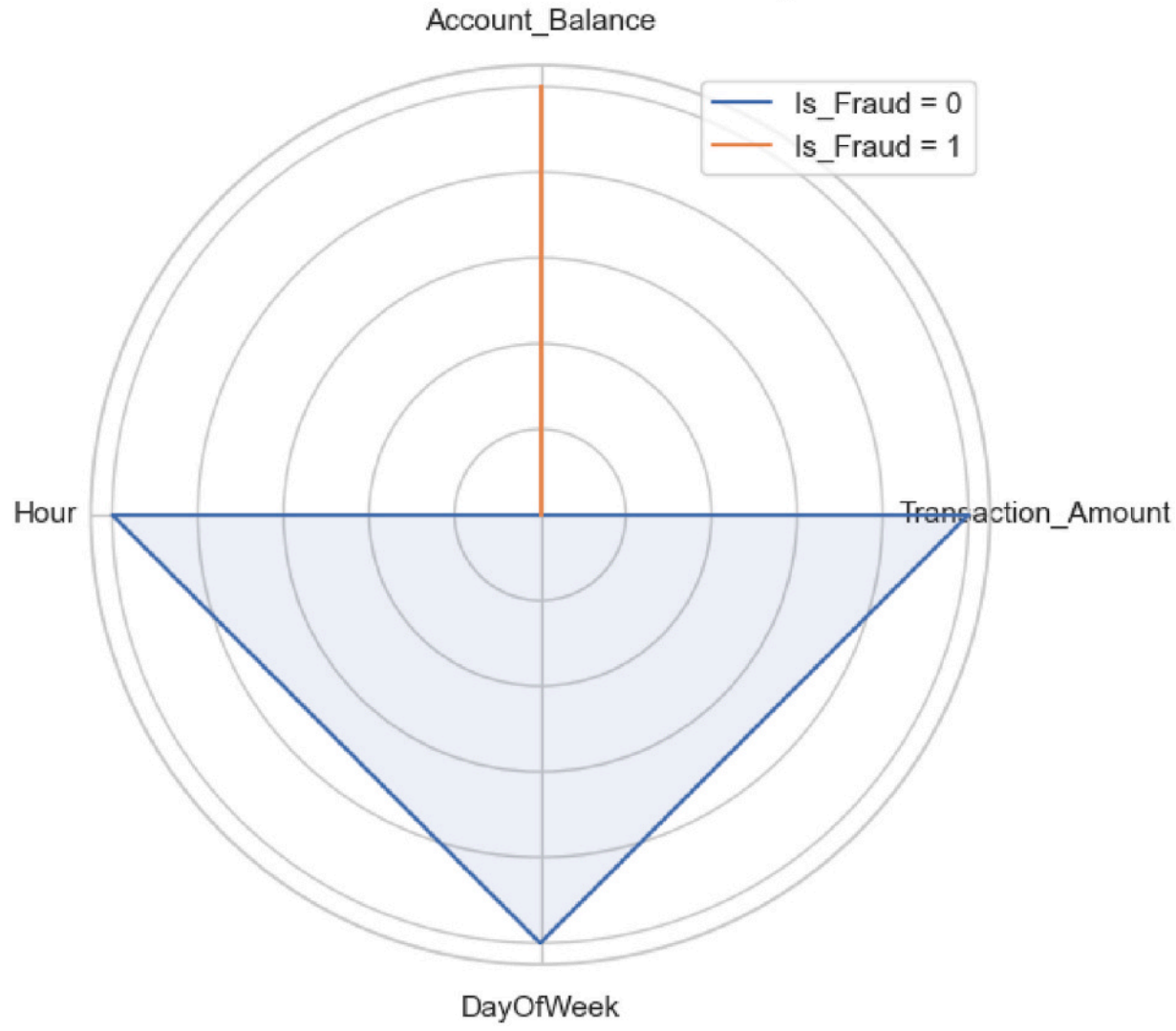

**Fig. 9.** Radar plot.

4. **Handles Both Categorical and Numerical Data**: TabNet can process **categorical features directly** (after simple label encoding) without requiring full one-hot encoding, which makes it **memory-efficient** and faster during training compared to MLPs that depend on large, sparse matrices.
5. Feature Selection for Contextual Decisions: Unlike models that assign a single global importance to features, TabNet's attention mechanism performs feature selection. This means it can prioritize unique features for different transactions, mimicking how a human analyst might focus on varying risk factors depending on the context of each specific case.
6. **Regularization through Sparse Feature Masks**: TabNet applies **sparsity constraints** during training, which helps avoid overfitting, especially important in fraud datasets where class distribution and behavior patterns can shift over time.
7. **Superior Performance in Imbalanced Settings**: Fraud datasets—even after SMOTE balancing—often have **subtle and non-linear class boundaries.** TabNet's decision-based architecture and attention mechanism make it better suited to **capture complex patterns** compared to vanilla FNNs or logistic regression.
8. **End-to-End Learning**: TabNet eliminates the need for ensemble methods (like stacking tree models with neural networks), providing a **unified and scalable deep learning approach** to fraud detection without the overhead of feature engineering pipelines.

TabNet is preferred over regular deep learning models for this bank fraud classification task because it combines high predictive performance with interpretability, efficiency, and a design tailored to the tabular nature of the dataset. Its ability to adaptively select features at each step and explain decisions makes it ideal for building reliable and explainable fraud detection systems in real-world banking environments.

### *4.5. Proposed TabNet methodological framework*

TabNet is a deep learning architecture specifically designed to handle tabular datasets, which are traditionally dominated by tree-based models like XGBoost and Random Forests. Developed by Google Cloud AI researchers, TabNet introduces a sequential attention mechanism that dynamically selects features at each decision step, enabling both interpretability and high predictive performance.

Unlike fully connected neural networks that treat all features equally, TabNet learns to focus on the most relevant features during training by using sparse feature masks. This allows the model to mimic decision tree behavior while maintaining the flexibility and scalability of neural networks.

Key advantages of TabNet include.

1. Interpretability through learned feature selection masks.
2. End-to-end learning with gradient descent, avoiding manual feature engineering.
3. Compatibility with GPU acceleration for large-scale datasets.
4. Superior performance on many tabular benchmarks compared to traditional ML methods.

TabNet is especially well-suited for fraud detection, credit scoring, healthcare predictions, and other structured data tasks where feature importance and decision transparency are crucial.

The proposed framework, illustrated in Fig. 10, presents a structured end-to-end workflow for robust fraud detection using TabNet. The process begins with the Data Processing Pipeline, where the bank fraud dataset is loaded, categorical features are encoded, and data is cleaned. Exploratory Data Analysis (EDA) is performed to understand the underlying structure, followed by class balancing using SMOTE to handle label imbalance. This pre-processed data is then subjected to a 3-Fold Cross-Validation Pipeline, where different combinations of hyperparameters are evaluated across three training-validation splits to identify the most suitable configuration. Once the optimal hyperparameters are finalized, a full training of the TabNet model is conducted using the entire dataset. The trained model's performance is then assessed through evaluation metrics like ROC-AUC and a confusion matrix. Finally, in the Evaluation and Comparison Pipeline, the results are benchmarked against other deep learning models such as LSTM, GRU, CNN1D, and DNN ensuring the effectiveness and reliability of the proposed TabNet architecture in detecting fraudulent transactions. This modular, iterative design ensures both generalizability and model interpretability in real-world financial fraud scenarios.

The algorithm for the overall flow is as shown in Algorithm 1.

**Algorithm 1.** Bank Fraud Classification using Cross-Validation and Deep Learning

**Require:** Dataset D with features X and labels y belong to $(0, 1)$; number of folds K
**Ensure:** Trained model and evaluation metrics

1: Encode categorical features in X
2: Normalize or scale numerical features if required
3: Convert X, y into numerical format (NumPy or tensor)
4: Initialize $K$-Fold Stratified Cross-Validation on D
5: **for** each fold i = 1 to K **do**
6: Split D into training set $T_i$ and validation set $V_i$
7: Initialize deep learning model $M_i$ (e.g., TabNet, MLP)
8: Train $M_i$ on $T_i$ with:
- Epochs = 30, batch size, optimizer (e.g., Adam)
- Early stopping with validation AUC and patience

9: Predict on $V_i$ to obtain $y\wedge_i$
10: Compute and store metrics (ROC-AUC, F1, Confusion Matrix)
11: **end for**
12: Compute means and standard deviations of all metrics across K folds
13: Retrain final model $M_{\text{final}}$ on full dataset D with best parameters
14: Predict on D and export:
- Predictions with probabilities
- Feature importance (if available)
- ROC curve and confusion matrix

15: **return** $M_{\text{final}}$ and metrics

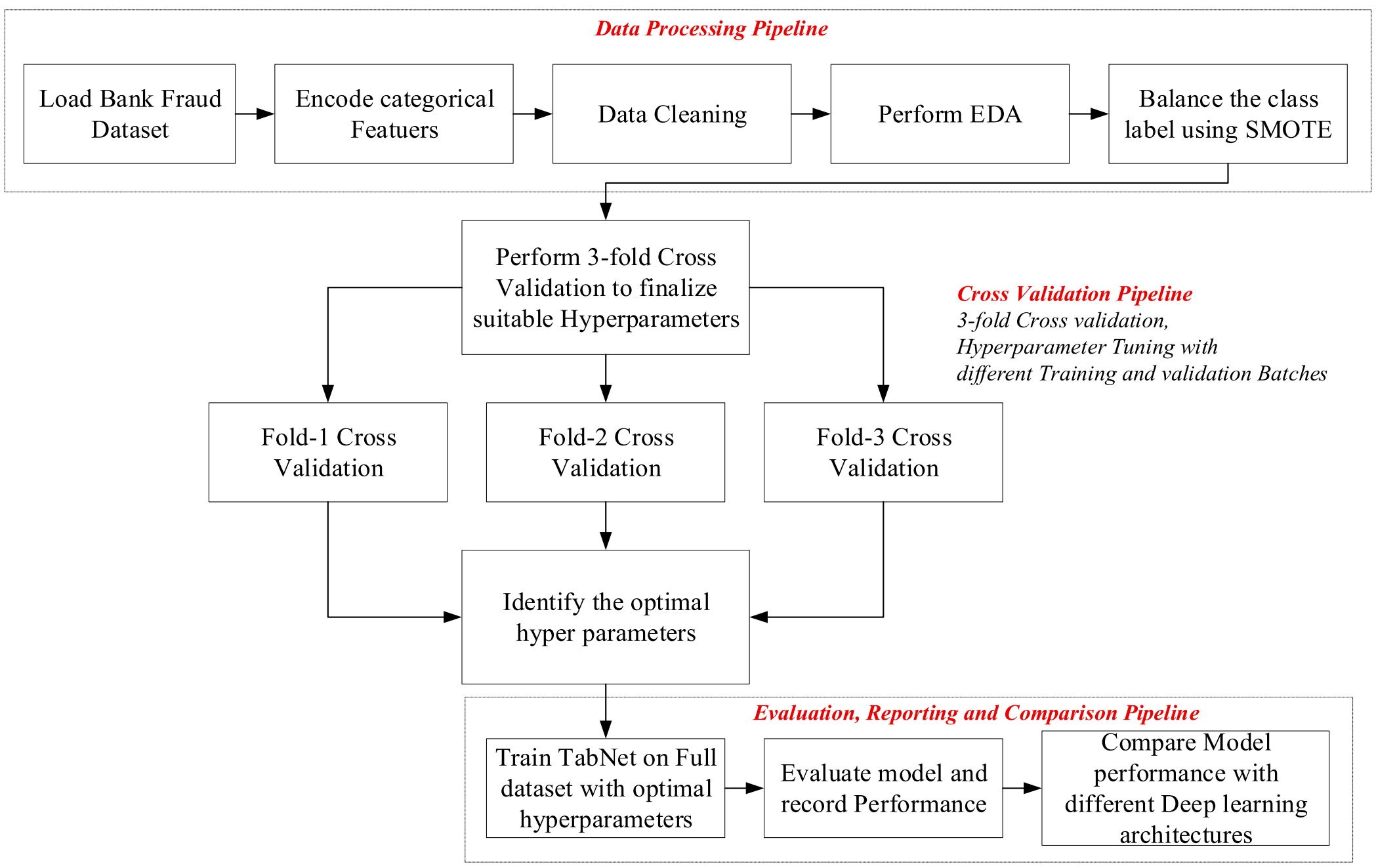

**Fig. 10.** TabNet-based cross-validation framework for fraud detection architecture.

## 5. Experiment results, analysis, and discussion

The TabNet model designed in this experiment has the parameters as shown in Table 4.

### 5.1. Experimental setup

The experimentation is carried out on a Ryzen 7 Nitro Machine with an Nvidia RTX 3050 graphics card with 16 GB RAM and a 6 GB cuda enabled device. The implementation was done using PyTorch, a popular deep learning framework using Python.

### 5.2. Results and analysis

In this section, we evaluate and compare the performance of various deep learning architectures for the task of detecting bank fraud. The models evaluated include LSTM, GRU, CNN-1D, TabNet, and a baseline Deep Neural Network (DNN). The evaluation was conducted using a 3-fold cross-validation and final training on the full dataset. Key metrics considered include ROC-AUC and overall accuracy. We now examine the performance of each model in detail.

The results obtained from the DNN model reveal significant limitations in its ability to distinguish between fraudulent and non-fraudulent transactions. As shown in the confusion matrix in Fig. 11a, all samples from both classes (62,248 labelled as "Not Fraud" and 62,248 labelled as "Fraud") were predicted as "Fraud." This indicates a complete classification collapse, where the model defaults to predicting a single class regardless of the input features. Consequently, the ROC-AUC score is 0.5, depicted in Fig. 11b, which is equivalent to random guessing and suggests that the model lacks any discriminatory power. This is further confirmed by the ROC curve, which lies along the diagonal and does not show any separation capability between the classes. The overall accuracy also stands at 0.5, which again aligns with random prediction behavior on a balanced dataset.

The CNN1D-based classifier exhibited suboptimal performance on the bank fraud detection task. The 3-fold cross-validation yielded a mean ROC-AUC score of 0.5005, and the model's overall classification accuracy was 50 %, suggesting performance equivalent to random guessing. The confusion matrix shown in Fig. 12a indicates that the model predicted all instances as the positive class (“Fraud”), resulting in no true negatives or true positives, which leads to extreme class imbalance in predictions. This misclassification pattern is further reflected in the ROC curve shown in Fig. 12b, which closely follows the diagonal, confirming a lack of discriminative capability. These results imply that the CNN1D architecture, in this case, failed to capture meaningful patterns in the tabular fraud dataset, likely due to the lack of spatial structure that CNNs typically leverage in image data. Thus, CNN1D may not be a suitable architecture for this domain without significant feature engineering or architectural adjustments.

The GRU-based model demonstrated limited effectiveness in detecting fraudulent transactions. With a mean ROC-AUC of 0.517 as shown in Fig. 13b during 3-fold cross-validation, and an overall validation accuracy of 50 %, its predictive power was only marginally better than random guessing. When evaluated on the full dataset using the tuned hyperparameters, the confusion matrix shown in Fig. 13a revealed a sizeable number of both false positives (26,485) and false negatives (34,473), highlighting challenges in class discrimination. The ROC curve further supported these findings, showing a nearly diagonal trend with a final ROC-AUC of 0.517. These results suggest that while the GRU architecture has some capacity to learn from the data, it lacks sufficient discrimination power for effective fraud classification in its current configuration.

The Long Short-Term Memory (LSTM) model yielded a mean ROC-AUC score, as shown in Fig. 14b, of 0.5118 over three cross-validation folds, indicating a slightly better-than-random performance in distinguishing between fraudulent and non-fraudulent transactions. The overall accuracy was 0.50, suggesting that the model struggled to generalize effectively on the balanced dataset. The confusion matrix shown in Fig. 14a reveals significant misclassifications: the model predicted 51,559 true negatives and 11,168 true positives, while also producing 10,689 false positives and 51,080 false negatives. The ROC curve confirms this marginal separation capability, hovering close to the diagonal reference line. This implies that although LSTM is designed to model sequential patterns, it may not effectively capture distinguishing temporal features in this static tabular fraud dataset, likely due to the absence of temporal dependencies or overfitting challenges.

The TabNet model demonstrated exceptional performance in the fraud detection task. It achieved a mean ROC-AUC score of 0.9739, as shown in Fig. 15b, over 3-fold cross-validation and an overall accuracy of 97.39 % with the best hyperparameters. The confusion matrix, as shown in Fig. 15a, indicates that the model correctly classified 62,248 non-fraud cases and 58,775 fraud cases, with only 3473 fraud cases misclassified as non-fraud cases and no false positives for the non-fraud class. This elevated level of precision and recall is further reinforced by the ROC curve, which remains close to the top-left corner, indicating excellent discriminatory power. TabNet's ability to handle tabular data with feature selection at each decision step appears to significantly outperform other deep learning models in this use case.

**Table 4**
TabNet model parameters.

| Parameter | Value |
|---|---|
| Model Type | TabNetClassifier |
| Max Epochs | 100 |
| Batch Size | 1024 |
| Virtual Batch Size | 128 |
| Eval Metric | ROC-AUC, Accuracy |
| Cross-validation | StratifiedKFold (5) |
| Interpretability | Feature Importance |

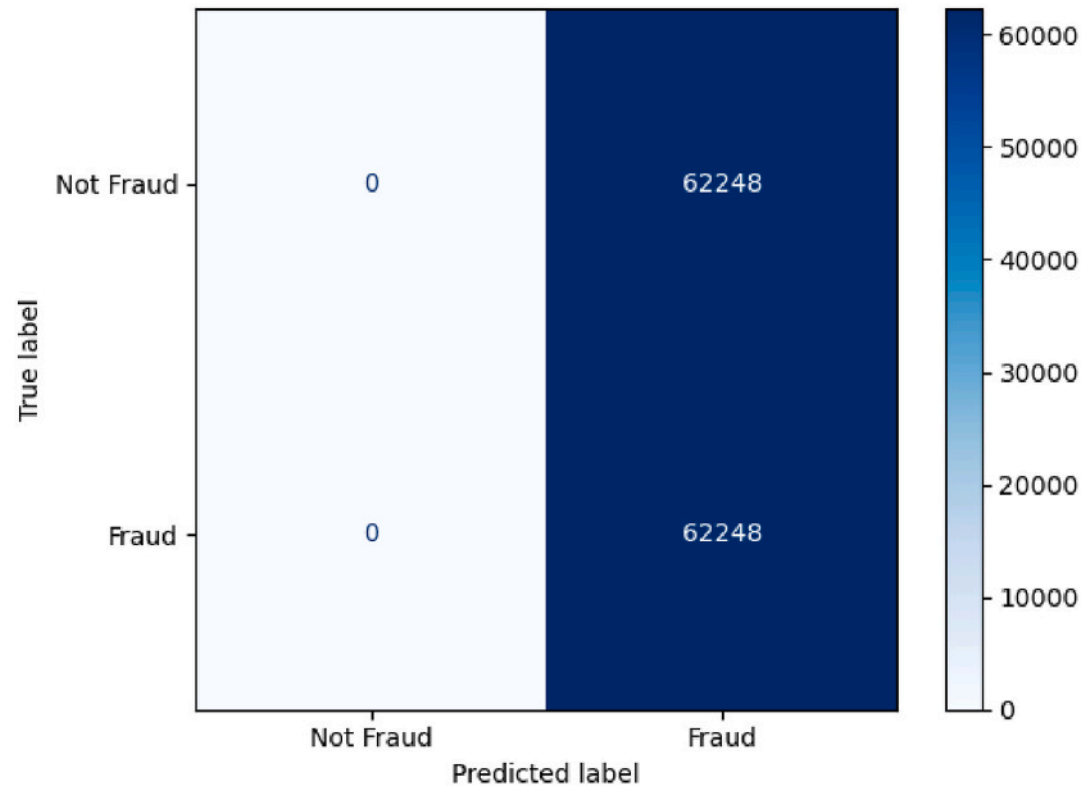

**Fig. 11a.** Confusion Matrix of DNN with 3-fold cross-validation.

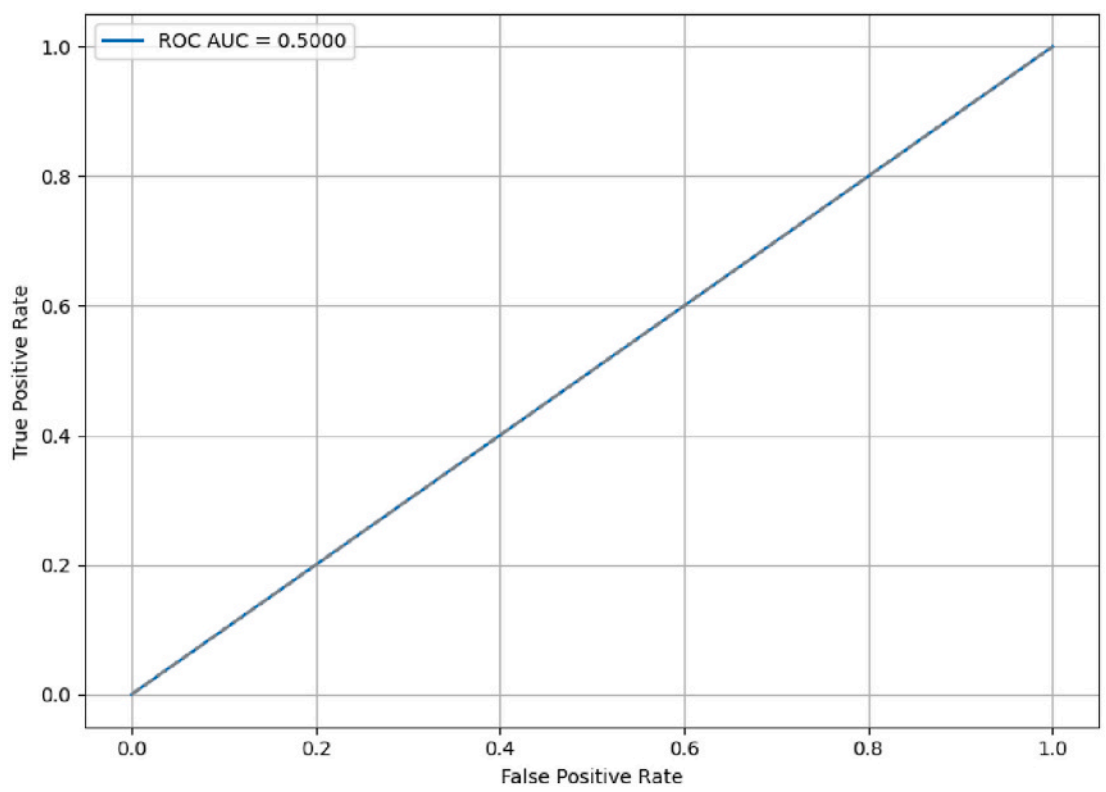

**Fig. 11b.** ROC curve of DNN with 3-fold cross-validation.

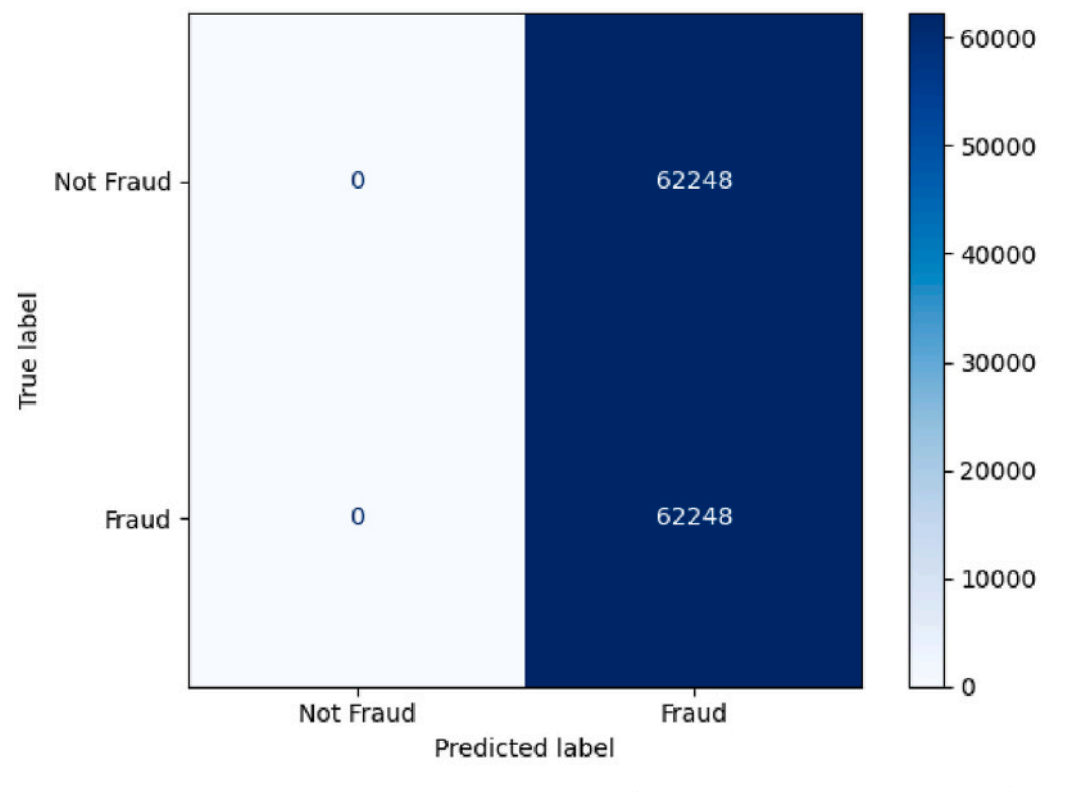

**Fig. 12a.** Confusion Matrix of CNN-1D with 3-fold cross-validation.

Table 5 summarises the comparative performance of alternative deep neural network models.

### 5.3. Key points of the experiment results

Key Points: The following are the key takeaways resulting from this experimentation:

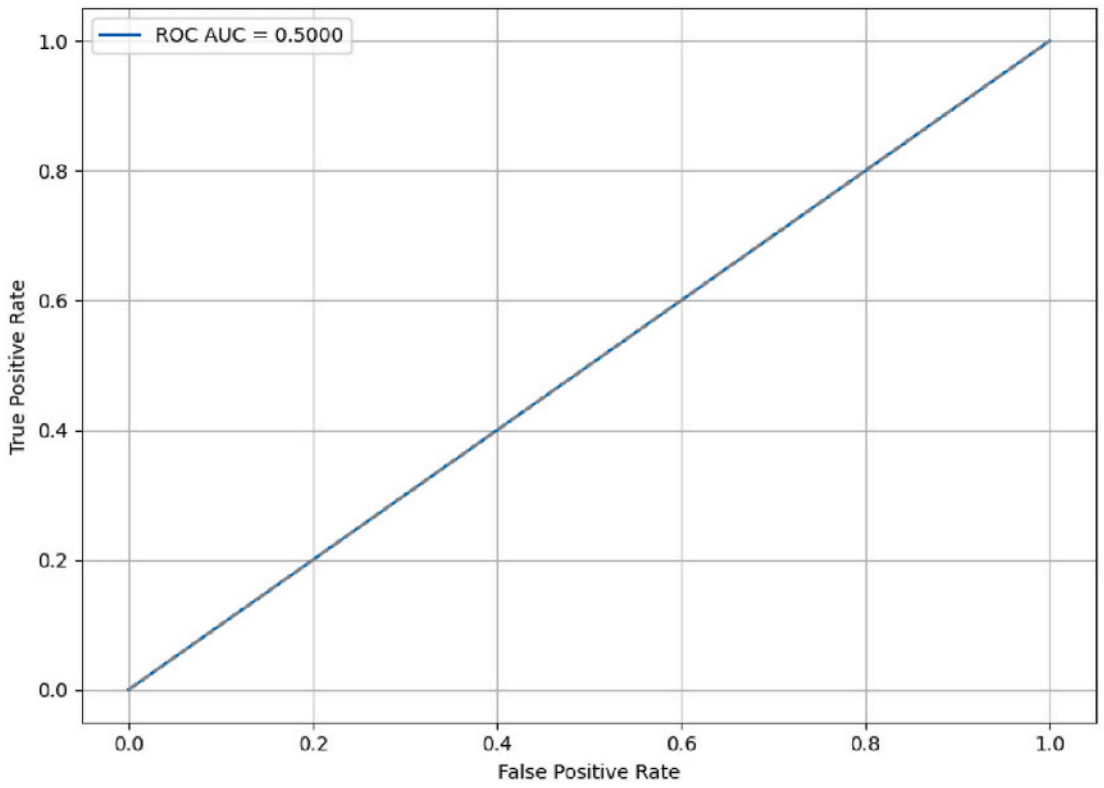

**Fig. 12b.** ROC curve of CNN-1D with 3-fold cross-validation.

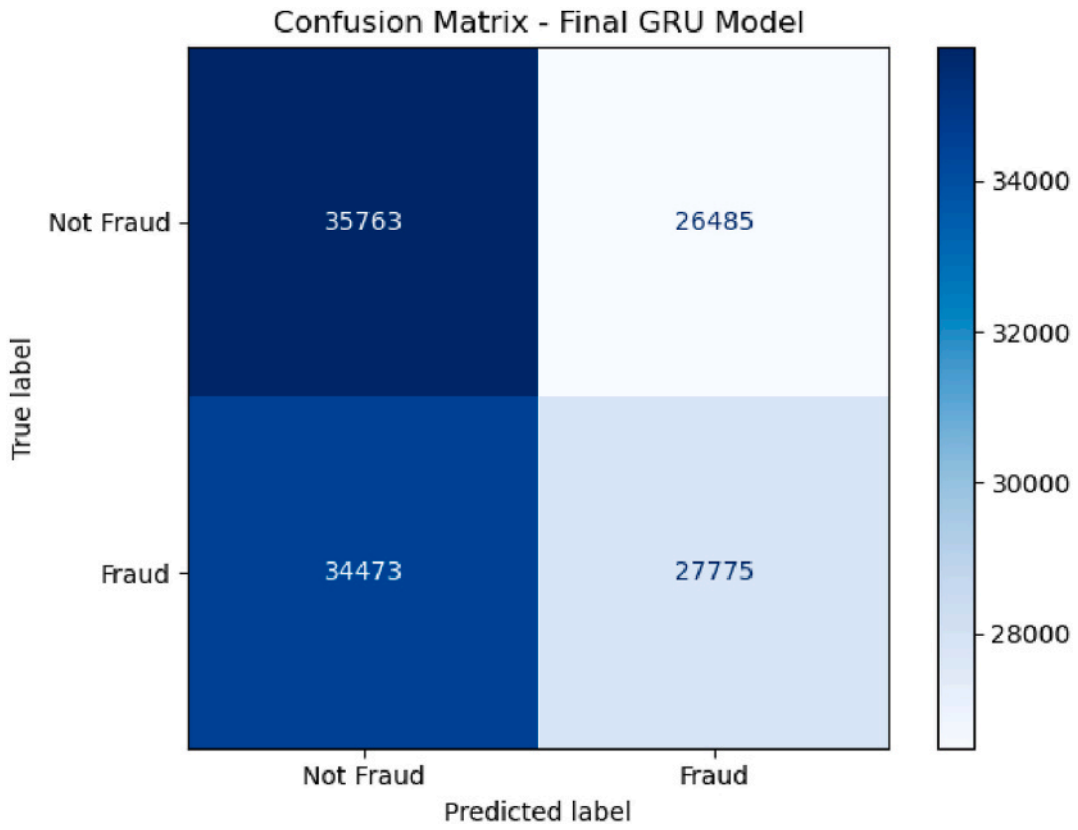

**Fig. 13a.** Confusion Matrix of GRU with 3-fold cross-validation.

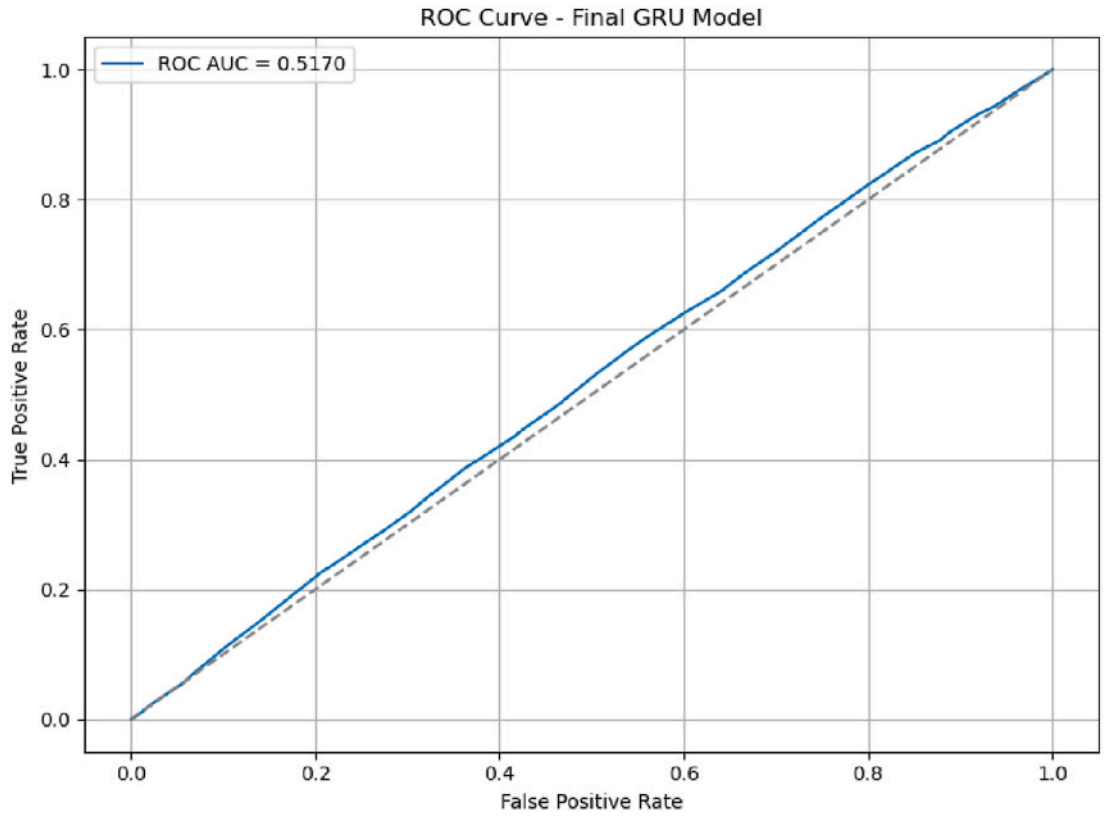

**Fig. 13b.** ROC curve of GRU with 3-fold cross-validation.

### *5.3.1. Evidence from EDA-based insights*

Important behavioral patterns were provided by the EDA, which both informed and confirmed the model findings

- By Time of Day: Between 8 a.m. and 5 p.m., with a peak between 10 and 11 a.m., many transactions—including fraudulent ones—took place.
- Contrary to conventional wisdom, fraud does not appear to cluster during off-hours, casting doubt on the notion that irregular hours necessarily indicate danger.

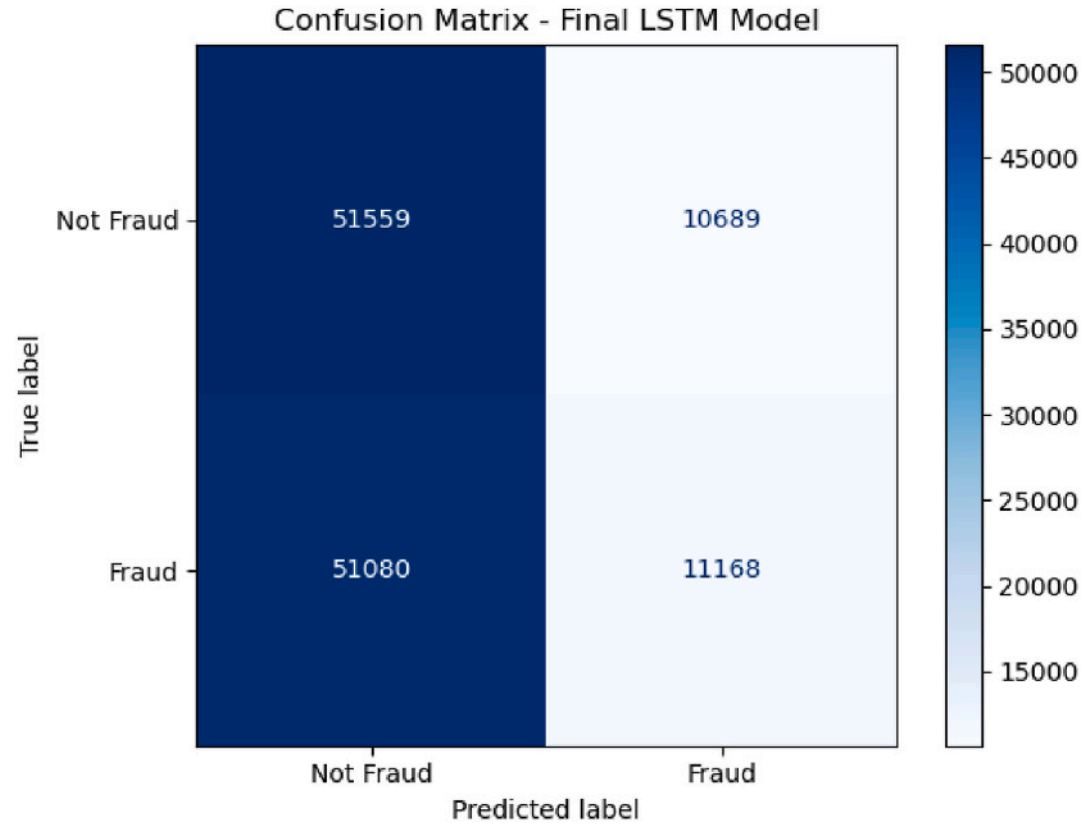

**Fig. 14a.** Confusion Matrix of LSTM with 3-fold cross-validation.

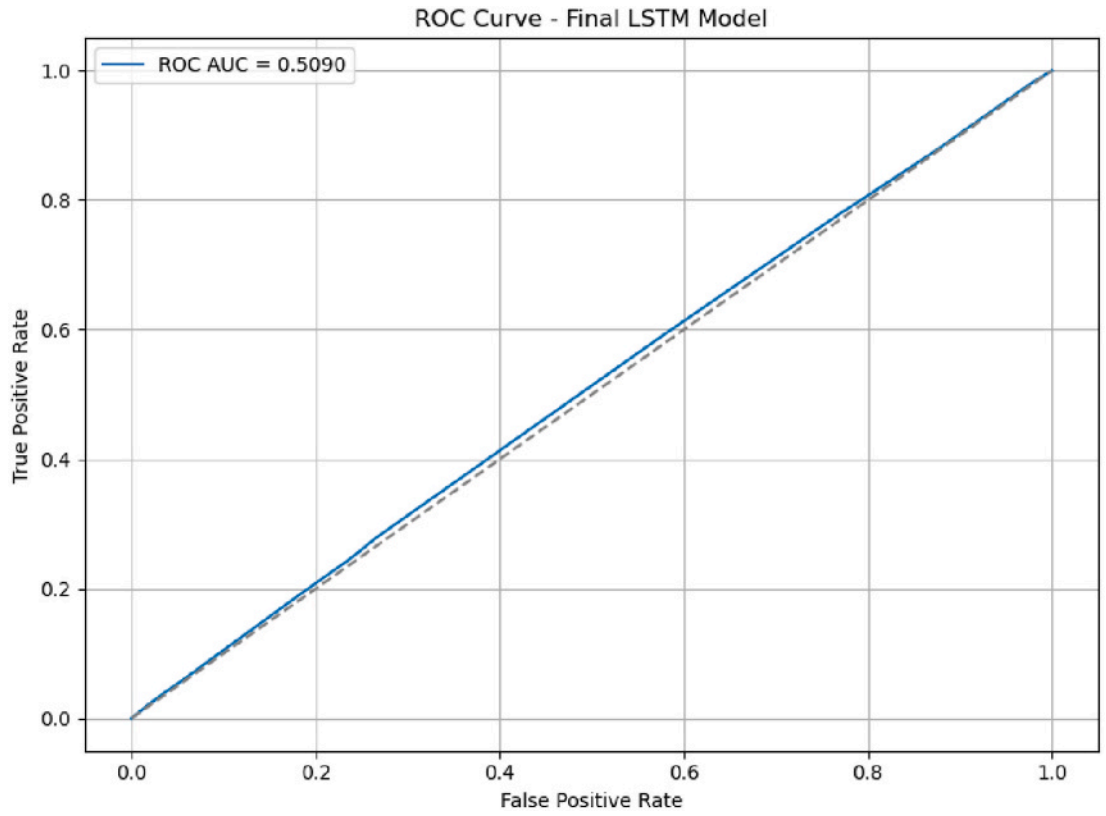

**Fig. 14b.** ROC curve of LSTM with 3-fold cross-validation.

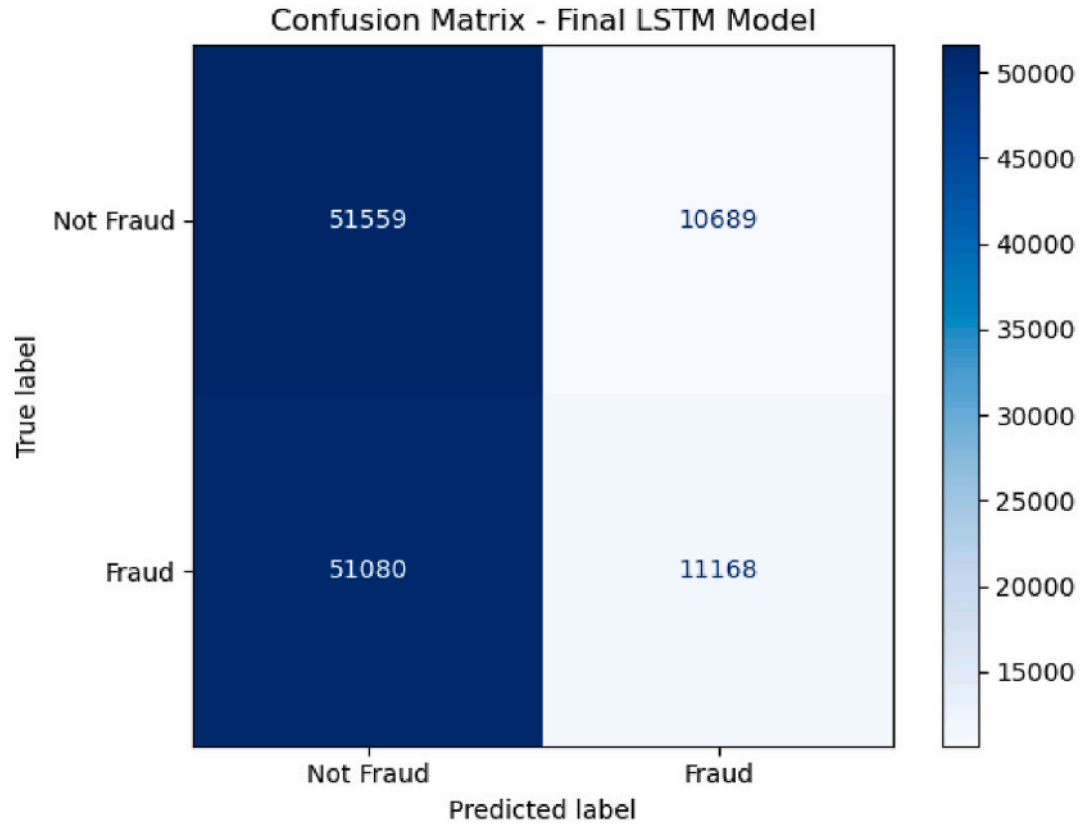

**Fig. 15a.** Confusion Matrix of TabNet with 3-fold cross-validation.

- Context: Criminals may purposefully imitate legal behaviors to evade filters that are based on rules.

*5.3.2. Patterns by day of the week*

Mid-week transactional overload is used by fraudsters to conceal irregularities, as fraud peaks on Wednesdays and Thursdays, which coincide with the busiest times of the week for banks and e-commerce platforms.

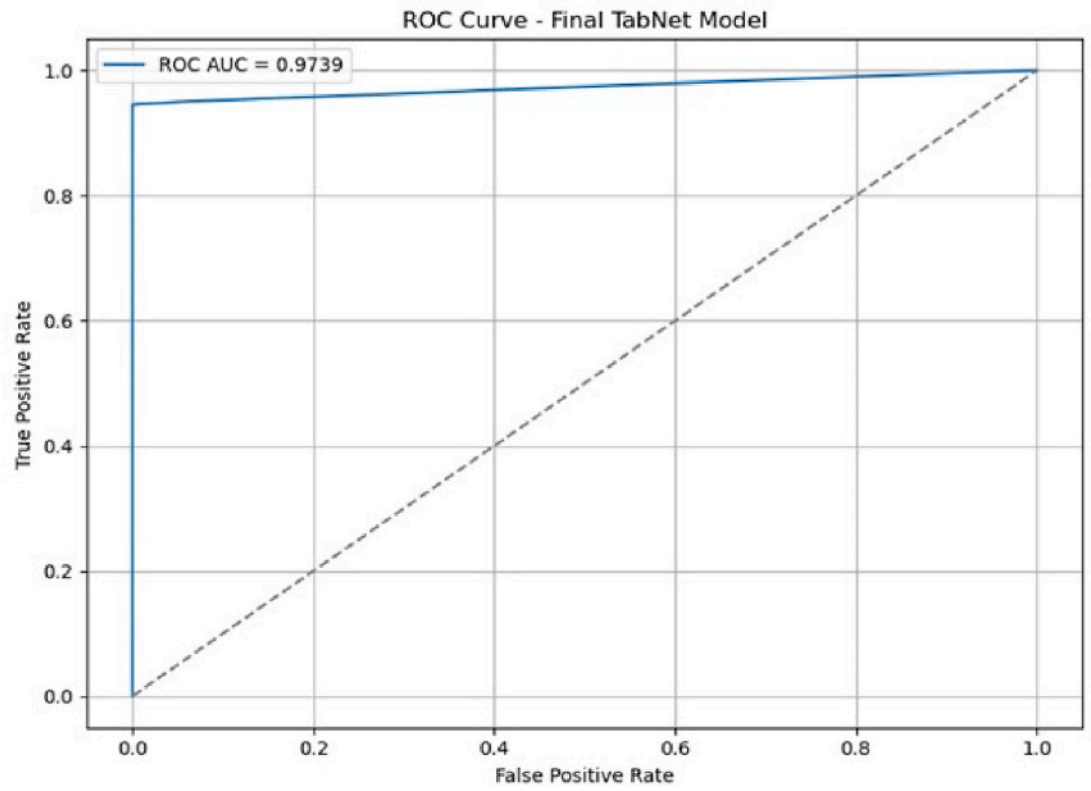

**Fig. 15b.** ROC curve of TabNet with 3-fold cross-validation.

**Table 5**
Comparative performance of alternative deep neural network models.

| Models | Cross-Validation Accuracy | | | Cross-Validation AUC | | | Full Set Accuracy (1) | Full set AUC (1) |
|---|---|---|---|---|---|---|---|---|
| | Fold-1 | Fold-2 | Fold-3 | Fold-1 | Fold-2 | Fold-3 | | |
| TabNet | 97 % | 97 % | 97 % | 0.9736 | 0.9733 | 0.9747 | 97 % | 97 % |
| LSTM | 50 % | 51 % | 51 % | 0.5134 | 0.5078 | 0.514 | 50 % | 51 % |
| GRU | 51 % | 52 % | 51 % | 0.5142 | 0.5211 | 0.5149 | 51 % | 52 % |
| CNN-1D | 49 % | 50 % | 50 % | 0.49 | 0.506 | 0.5002 | 50 % | 50 % |
| DNN | 50 % | 50 % | 50 % | 0.5 | 0.5 | 0.5 | 50 % | 50 % |

#### *5.3.3. The account balance and amount of the transaction*

- Account balances were greater in fraudulent transactions despite smaller transaction amounts.
- The transaction patterns of legitimate users were more spread.

According to behavioral interpretation, fraudsters may attempt "micro-fraud" by making small withdrawals from accounts with large balances to evade detection.

#### *5.3.4. A graph and heatmap for radar data*

o trait combinations are better predictors of fraud than any one trait, according to the visualizations.
o On weekday mornings and early afternoons, the heatmap showed that fraudulent activity was at its peak.

#### *5.3.5. The necessity of attention-based models, such as TabNet, which may dynamically prioritize feature combinations, is justified by the model's significance*

- TabNet emerged as the best-performing model, achieving an impressive ROC-AUC of 0.9739 and accuracy of 97.39 %, with highly accurate predictions across both fraud and non-fraud classes. It demonstrated strong generalization and robust class separation, making it highly suitable for tabular fraud detection tasks.
- LSTM and GRU models performed marginally better than random guessing, with ROC-AUC scores of 0.5118 and 0.517, respectively. However, both models showed significant misclassifications and confusion matrix imbalances, suggesting that sequence-based architectures may not be well-suited for this non-sequential, tabular dataset.
- CNN1D and DNN models completely failed to distinguish between classes, with ROC-AUC scores close to 0.5 and accuracy of 50 %, suggesting behavior equivalent to random prediction.
- Memory constraints also favor models like TabNet in practical deployment. Deep sequential models such as LSTM and GRU can be resource-intensive, especially when trained on large-scale transactional data, which may limit their use in real-time systems. TabNet's lighter inference footprint and built-in feature selection make it a more deployable and efficient choice for real-world bank fraud detection systems.

Discussion: Using structured, imbalanced tabular data for the objective of identifying fraudulent bank transactions, the experimental results show that different deep learning architectures perform significantly differently in the model performance evaluation. In this research, we will examine how to interpret these findings in relation to XAI, operational risk management, and fraud detection more broadly. With the help of exploratory data insights and learning theory, we focus on how the TabNet architecture outperforms competing models such as DNN, CNN1D, GRU, and LSTM.

### 5.4. Discussion of experimental results

Performance Comparison Across Models:

#### 5.4.1. TabNet outperformance

Among the evaluated architectures, the TabNet model outperformed the others with:

- AUC-ROC: 0.9739
- Precision: 97.39 percent
- Resolved all legitimate transactions without a single false positive.
- A high rate of accurate detections for instances of fraud

According to these measures, TabNet is the best at detecting interactions between high-order features in tabular data. The ROC curve clung to the upper left corner, suggesting outstanding class separability, and the confusion matrix revealed few misclassifications. Featuring dynamic attention and interpretability through feature masks, it is specifically designed for structured financial data and meets the requirements of explainable fraud detection systems set by regulators.

#### 5.4.2. Long Short-Term Memory (LSTM) and general recurrent units (GRU)

- LSTM (ROC-AUC 0.5118) and GRU (ROC-AUC 0.517)
- The lack of temporal connections in tabular transactional data makes it difficult for these systems to derive meaningful representations, as they are designed for sequential data. The confusion matrices showed a high rate of false positives and negatives, indicating that the classes were not well segmented.
- They have not been performing up to expectations because of
- Transactions are not always consecutive for each user, and the dataset does not contain any explicit temporal sequences.
- The possibility of overfitting due to insufficient data sequences and lengthy training durations.
- Recurrent layers are not as helpful because there is not enough temporal granularity.

#### 5.4.3. CNN1D and DNN: classification collapse

The CNN1D and DNN models were both severely underwhelmed:

- ROC-AUC close to 0.5, which is the same as taking a wild guess.
- Despite using SMOTE for balancing class, accuracy remains at 50 %.
- Because of declaring all cases as "fraud," the classification system broke down.

Specifically, the DNN resorted to mode bias (class collapse) instead of learning discriminatory features. Inadequate tuning of the architecture for imbalanced data, diminishing gradients due to depth, and poor weight initialization are all potential causes.

The flat form of the bank transaction dataset was not well-suited for CNN1D, which is normally good at detecting spatial patterns in image or sequence data. Contrary to picture pixels, tabular characteristics do not contain any "locality" that might potentially damage CNN's convolutional filters.

### 5.5. Summary of key findings

| Model | AUC | Accuracy | Strengths | Limitations |
|---|---|---|---|---|
| TabNet | 0.9739 | 97.39 % | High interpretability, low errors | May miss subtle new fraud patterns |
| LSTM | 0.5118 | 50–51 % | Slight sequential awareness | Poor tabular, static data |
| GRU | 0.517 | 51–52 % | Faster training than LSTM | High misclassification |
| CNN1D | 0.50 | 50 % | None in this context | Not suitable for tabular data |
| DNN | 0.5 | 50 % | Simple to implement | Collapsed predictions, no separation |

## 6. Conclusion

### 6.1. Stressing once again how critical it is to avoid financial fraud in the modern digital age

Both the convenience and the risk within global financial ecosystems have been magnified by the exponential expansion of digital financial transactions, driven by fintech innovation and increasing mobile banking usage. The Indian financial sector is especially vulnerable to fraud due to the country's large unbanked population, shift toward mobile-first banking, and high rate of digital onboarding. Due to their high false positive rates, limited feature generalization, and inflexible logic, traditional rule-based algorithms have failed to capture the dynamic patterns of fraud.

This study aimed to address a relevant and important question in this context: Is it possible to use adaptive deep learning models, particularly TabNet, to detect emerging patterns of complex bank fraud using structured and imbalanced transactional data more effectively than traditional deep neural networks?

### *6.2. An overview of the Study's goals and methods*

Three primary goals motivated the current investigation:

a) To enhance the operational risk framework for detecting fraud in financial institutions by developing and evaluating a predictive system based on deep learning.
b) To discover which neural architectures (DNN, GRU, LSTM, CNN1D) are most suited for structured fraud data by evaluating model performance.
c) Create and test TabNet, a novel architecture that uses sparse attention techniques to determine which features are relevant, allowing for accurate fraud detection, generalizability, and interpretability.
d) To improve the transparency and interpretability of the fraud detection process by identifying and visualizing the key predictive features and decision patterns used by the models.

- We were able to accomplish this by utilizing a structured bank transaction dataset derived from areal-life and India-centric Kaggle competition, which included an extremely skewed fraud class.
- After using SMOTE to level the playing field, we dove headfirst into sophisticated exploratory data analysis (EDA) to sniff out suspicious patterns of transactions and time that could point to fraud.
- We used 3-fold cross-validation to train and assess five different deep learning models.
- The use of confusion matrices, accuracy, and ROC-AUC assessed their effectiveness.
- The use of visual tools such as radar plots, heatmaps, and temporal frequency distributions provided further support for analytical findings.

### *6.3. Important results from experiment*

a. TabNet: An Outstanding Example

The most effective model that was examined was TabNet, which achieved

- A ROC-AUC value of 0.9739
– 97.39 percent accuracy in classification
- Minimal instances of both false positives and negatives
- The model met the needs for transparency and regulatory alignment in high-risk domains like fraud detection with its built-in interpretability (via feature masks) and sparse attention mechanism, which allowed it to dynamically prioritize key features during training.

b. Conventional DNN, LSTM, GRU, and CNN1D Are Not Sufficient for Tabular Fraud Data
  - Every other model showed deficient performance:
    o LSTM and GRU: A little better than randomness, but high misclassification.
    o CNN1D: Designed for spatial/temporal data; inappropriate for tabular input.
    o DNN: collapsed predictions; ROC-AUC 0.5

When applied to structured, nonsequential data, these models were unable to detect fraud because they overfit the minority class or did not detect complicated feature interactions.

### *6.4. Theoretical contribution*

The study contributes to the ongoing literature on financial fraud detection and deep learning by:

- **Demonstrating the limits of generic neural architectures** (DNNs, LSTMs, GRUs) for tabular data fraud detection.
- **Establishing TabNet as a theoretically sound and empirically validated alternative**, particularly in structured, high-dimensional, and imbalanced financial datasets.
- **Reinforcing the importance of interpretable AI (XAI)** in critical domains, aligning model outcomes with compliance and accountability frameworks like Basel III, AML regulations, and RBI fraud guidelines.

### 6.5. Practical and operational contribution

On the applied side, the study has multiple implications:

- **Banking and Fintech Industry**: TabNet provides a scalable and interpretable fraud detection pipeline for digital transaction systems in India and beyond.
- **Risk Management and Auditing**: With near-perfect precision and low false alarms, TabNet supports effective early warning systems and internal control audits.
- **Technology Stack Guidance**: Our experimental setup and pipeline (data ingestion → SMOTE balancing → TabNet cross-validation → interpretability layer) can serve as a blueprint for financial fraud detection systems across institutions.

### 6.6. Insights from exploratory data analysis

A noteworthy aspect of this research is the synergy between model outcomes and insights derived from EDA:

- **Temporal patterns**: Fraudsters typically operate during business hours to blend in with legitimate traffic—contrary to the popular belief that fraud spikes during off-hours.
- **Behavioral fingerprints**: Fraudulent transactions often involve lesser amounts but are drawn from high-balance accounts.
- **Spatial and merchant category patterns**: Cross-linking location and category data could further enhance detection, a promising area for future research.

This behavioral profiling, when integrated into model architecture through engineered features or attention mechanisms (as TabNet does), drastically improves fraud detection accuracy.

### 6.7. Addressing the research gap

The research fills several gaps in literature and practice:

- **Few studies benchmark TabNet** against other DNN variants in a real-world context of fraud detection.
- **India-specific financial datasets** are underrepresented in the literature; this study utilizes an Indian dataset to reflect regional patterns of digital banking fraud.
- **Most fraud detection models ignore interpretability**—we integrate XAI through TabNet's intrinsic design.

### 6.8. Model robustness and deployment readiness

The findings also confirm TabNet's:

- **Robustness to imbalanced data** (even post-SMOTE)
- **Compatibility with GPU acceleration** for real-time use cases
- **Modular design**, making it amenable to integration with APIs, cloud-based fraud analytics systems, or on-premises banking core solutions.

Its relatively lightweight inference footprint ensures that it can be deployed in both cloud and edge environments, making it suitable for use in real-time mobile banking apps and point-of-sale systems.

### 6.9. Policy implications

For policymakers, financial regulators, and banking risk governance boards, this research offers actionable evidence:

- **Fraud detection systems must evolve from static to adaptive** and from opaque to explainable.
- **Compliance with data protection and auditability** must be built into AI models from design to deployment.
- **The cross-sector adoption of interpretable AI**, especially in sectors with high exposure to operational and reputational risk (e.g., insurance, retail lending, microfinance), is crucial.

### 6.10. Recommendations for practitioners

a. **Replace legacy rule-based fraud systems** with adaptive models, such as TabNet, that combine precision with interpretability.
b. **Incorporate real-time behavioral analytics** (time-of-day, merchant profiling, device patterning) into transaction scoring algorithms.
c. **Establish fraud pattern heatmaps and dashboards** for operational teams based on TabNet's feature importance outputs.

d. **Develop ensemble frameworks** that integrate TabNet with anomaly detection modules (e.g., Isolation Forest) for broader fraud coverage.

#### 6.11. *Limitations of the study*

While the study demonstrates promising results, a few limitations must be acknowledged:

- **Data limitation**: The dataset, though robust, is India-specific. Global generalization requires multi-regional datasets and multi-lingual feature normalization.
- **Real-time integration not assessed**: While the models were assessed offline, their latency and throughput in live environments were not evaluated.
- **Concept drift**: Fraud patterns evolve. Though TabNet can retrain efficiently, the study does not explore online learning strategies or continual adaptation frameworks.

These limitations highlight avenues for future research, including real-time pipeline testing, federated learning deployment, and assessments of adversarial robustness.

#### 6.12. *Directions for future research*

- **Concept Drift and Lifelong Learning**: Implement online TabNet or integrate with continual learning algorithms to adapt to new fraud patterns in real-time.
- **Hybrid Architecture**: Combine TabNet with graph neural networks (GNNs) or transformer models to model relationships between customers, merchants, and devices.
- **Cross-domain Transfer Learning**: Train models on financial datasets and evaluate them on insurance or e-commerce fraud to assess generalizability.
- **Explainability Enhancements**: Integrate TabNet feature importance outputs with SHAP or LIME for enhanced interpretability of multi-layer models.

#### 6.13. *Final reflection*

This study confirms that **the future of fraud detection lies in adaptive, interpretable, and tabular-data-native deep learning architectures**. In a field where accuracy must be balanced with auditability and real-world constraints, TabNet emerges as a front-runner capable of meeting regulatory, operational, and technological expectations.

Our work advances both scholarship and practice by demonstrating how intelligent architecture choice—backed by empirical rigor and behavioral insights—can deliver secure, scalable, and interpretable fraud detection systems. It is a timely contribution to India's fast-evolving financial AI landscape and a compelling model for global adaptation.

#### 6.14. *Study implications*

##### 6.14.1. *Implications for theory*

a. The use of sparse attention in TabNet's dynamic feature selection is a key component to its performance. While DNNs treat all features the same, TabNet:
  - Features that are important at each decision step are learned. Ignoring unnecessary input reduces noise. Feature masks provide interpretability.
  - Theoretically, this is in line with cognitive load, which states that reducing learning effort while increasing model generalization is achieved by prioritizing the most informative elements.

b. Explainability as Meeting Regulatory Requirements
  - Financial models used in the real world need to provide justifications for their actions, particularly in the following areas:
    o Dispute resolution (why was a transaction highlighted, for example?)
    o Reports on audits and compliance with AML and KYC regulations
  - TabNet is in line with current ideas of fintech governance because of its intrinsic interpretability, which answers the "black box" critique of neural networks.

c. Modeling with Aware of Risk

Financial institutions are required by Basel III and operational risk standards to implement fraud detection systems that:

- Reduce the occurrence of false positives, which impact consumer trust.
- Maximize the number of true positives, which limit financial damage.

According to our study findings, TabNet helps decrease both Type I and Type II errors, lending credence to optimization tactics that take risk into account when detecting fraud.

#### 6.14.2. *Relevance to real life*

a. Choosing a Model for Rollout is crucial to align data architectures, as shown by the failure of LSTM, GRU, DNN, and CNN1D.
b. The failure of LSTM, GRU, DNN, and CNN1D underlines the importance of data-architecture alignment. Tabular datasets with mixed categorical and numerical data are best overseen with models like:
   - TabNet
   - LightGBM
   - Explainable tree-ensemble hybrids

   Our study validates that using unsuitable architecture (e.g., CNN on tabular data) can lead to catastrophic misclassification.

c. According to TabNet's results, it could be useful for:
- Monitoring transactions in real time
- Creating fraud detection systems with dashboards that are easy to understand.
- Seamless incorporation with pre-existing fraud management platforms
- Its lightweight structure and inference speed also make it viable for edge computing or API-based fraud detection services.

d. A Plan for Feature Engineering
- The EDA and model results show that the properties of the devices and the time features (hour, day-of-the-week) are the most critical indicators of fraud.
- This ought to be kept for usage in the next modeling frameworks and enhanced with:
   - Behavioral fingerprints (user profiling based on frequency)
   - Cross-feature embeddings (for instance, city x merchant category)

e. Comparative Analysis with Past Literature
- Compared to earlier studies (e.g., Alarfaj et al., 2022; Alfaiz & Fati, 2022), our findings are consistent in two key aspects:
   o Deep models, such as LSTM and GRU, tend to underperform in static tabular settings.
   o Fraud detection is improved by **hybrid or adaptive models** capable of dealing with imbalanced, noisy data.
- However, our work adds novelty by:
   o Benchmarking **TabNet**, which is relatively underexplored in this domain.
   o Using a **fully Indian bank transaction dataset**, aligning results with local fraud patterns and operational contexts.

#### 6.14.3. *Explanation of TabNet's success in current use case*

TabNet's success can be attributed to:

- End-to-end learning: Unlike tree-based ensembles, TabNet learns the entire decision process without external feature selection.
- Sparse feature selection: Helps overcome the “curse of dimensionality” often encountered in high-dimensional financial datasets.
- Handling of SMOTE-balanced data: The architecture retains precision even when trained on synthetically balanced datasets.

This makes it uniquely positioned for banking applications where:

- Data is tabular and heterogeneous.
- Class imbalance is significant.
- Interpretability is non-negotiable.

#### 6.14.4. *Challenges and observed limitations*

Despite TabNet's success, our results suggest a few challenges:

- Even TabNet had **3473 false negatives**, suggesting that fraud patterns can still evade detection.
- Model performance may degrade if fraud strategies shift over time (concept drift), which is not assessed here.
- Data generalizability is limited to **Indian bank data**—performance on other datasets may differ.

## Declaration of interest

The authors declare that they have no known competing financial interests or personal relationships that could have appeared to influence the work reported in this paper.

## Acknowledgements

The authors gratefully acknowledge the support provided by Nitte Meenakshi Institute of Technology, Murdoch University, University of Dubai, Mohammed Bin Rashid School of Government, International Vaccine Institute, and IMC Krems University of Applied

Sciences.

## Data availability

Data will be made available on request.